%% file: bivariate.tex
\documentclass[10pt,journal,compsoc]{IEEEtran}

\ifCLASSOPTIONcompsoc
  \usepackage[nocompress]{cite}
\else
  \usepackage{cite}
\fi

\ifCLASSINFOpdf
   \usepackage[pdftex]{graphicx}
   \graphicspath{{../pdf/}{../jpeg/}}
   \DeclareGraphicsExtensions{.pdf,.jpeg,.png}
\else
   \usepackage[dvips]{graphicx}
   \graphicspath{{../eps/}}
   \DeclareGraphicsExtensions{.eps}
\fi

\usepackage{graphicx}
\usepackage{comment}
\usepackage{multirow} 
\usepackage[table,xcdraw]{xcolor}
\usepackage{caption}
\usepackage{subcaption}
\usepackage{epstopdf}
\usepackage{amsmath}
\usepackage{array}
\usepackage{tikz}
\usepackage[export]{adjustbox}
\usepackage{hyperref}
\usepackage{booktabs} 

\usepackage[finalnew]{trackchanges}

\begin{document}
\title{
\change{Bivariate Separable-Dimension Glyphs can Improve Visual Analysis of Holistic Features}{
Picturing Bivariate Separable-Features for Univariate Vector Magnitudes in Large-Magnitude-Range Quantum Physics Data}
}

\author{Henan~Zhao,~\IEEEmembership{Student Member,~IEEE,}
and
Jian~Chen,~\IEEEmembership{Member,~IEEE}

\IEEEcompsocitemizethanks{
\IEEEcompsocthanksitem H. Zhao is with the Department
of Computer Science and Electrical Engineering, University of Maryland, Baltimore County, MD 21250, USA. 
E-mail: henan1@umbc.edu.%
\IEEEcompsocthanksitem J. Chen is with the Department of Computer Science and Engineering,
The Ohio State University, Columbus, OH 43210, USA. 
E-mail: chen.8028@osu.edu.}
}

\markboth{Journal of \LaTeX\ Class Files,~Vol.~14, No.~8, August~2015}%
{Zhao \MakeLowercase{\textit{et al.}}: Bivariate Separable-Dimension Glyphs can Improve Visual Analysis of Holistic Features}

\IEEEtitleabstractindextext{%
\begin{abstract}

We present study results from two experiments to empirically validate  that separable bivariate pairs for univariate representations of large-magnitude-range vectors are more efficient than integral pairs.  The first experiment with 20 participants compared: one integral pair, three  separable pairs,  and one redundant pair, which is a mix of the integral and separable features. Participants performed three  \textit{local} tasks requiring
reading
numerical values, estimating ratio, and comparing two points.
\remove{, and looking for extreme values among a subset of points belonging to the same sub-group.}
The second 18-participant study compared three  separable pairs using three 
\textit{global} tasks
when participants must look at the entire  field to get an answer: find a specific target  in 20 seconds, find the maximum  magnitude in 20 seconds, and estimate the total number of vector exponents within 2 seconds. Our results also reveal the following: separable pairs led to the most accurate answers and the shortest  task execution time, while integral dimensions were among the least  accurate; it achieved high performance
only when a pop-out  separable feature (here  color) was added. To reconcile this finding with the existing literature, our second experiment suggests that the higher the separability, the higher the accuracy; the reason is probably that the emergent global scene created by the separable pairs reduces the subsequent search space.

\end{abstract}

}

\maketitle

\IEEEdisplaynontitleabstractindextext

\IEEEpeerreviewmaketitle

\IEEEraisesectionheading{\section{Introduction}\label{sec:introduction}}


\begin{figure*}[!t]
  \centering
\includegraphics[width=0.65\linewidth]{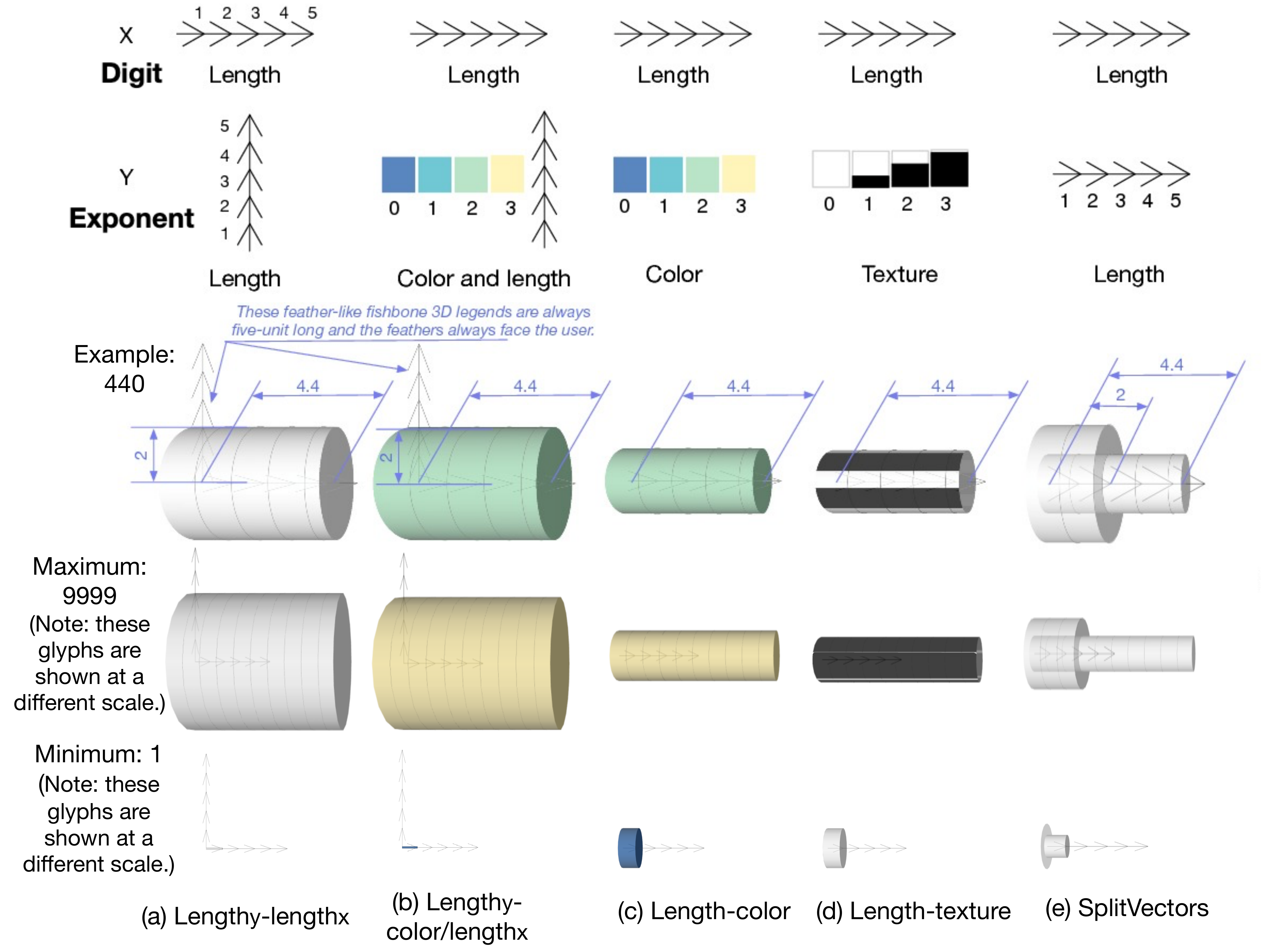}
\caption{
(a) Five bivariate configurations of univarate vector magnitude using the scientific notation. This example shows vector magnitude 440 ($4.4 \times 10^2$) with each depicted using two values: digit 4.4 and power 2. 
(b) Contours shown using the $length$-$color$ (LC) pair.
This work demonstrates that more separable pairs lead to efficient local comparisons of a couple of vectors. Global scene structures guided by more separable dimensions also led to more accurate and highly efficient strategies without the needs for synthesizing univariate magnitudes.
}
\label{fig:teaser}
\end{figure*}


\IEEEPARstart{B}{ivarate}
glyph  visualization is a common form of visual design in  which   a  dataset is  depicted by  two  visual variables, often   chosen   from  a  set  of  perceptually  independent graphical dimensions of shape, color, texture, size, orientation, curvature, and  so on{~\cite{fuchs2017systematic,ware2009quantitative}}. 
A bivariate glyph design has  been  used   to  show univariates for  quantum physicists at National Institute of Standards and Technology (NIST) to examine simulation results; thanks to their  team’s Nobel-prize-winning scalable  simulations, quantum physicists world-wide can now simulate at any scale. 
\add{A critical quantum-physics  analysis task is to understand spin (often  depicted as  vector)   magnitude  variations  because these   magnitudes  showing atom   behaviors  are  large in range  and  are often  not continuous where the magnitudes can vary  greatly in local regions.}
\remove{While a multitude of glyph techniques and design guidelines have been developed and compared in two-dimensions (2D){~\cite{fuchs2017systematic}~\cite{borgo2013glyph}~\cite{ward2008multivariate}}, a dearth of three-dimensional (3D)  glyph design principles exists. One reason  is that 3D glyph design is exceptionally
challenging because human judgments of metric 3D shapes and relationships contain large errors relative to the actual structure of the observed scene{~\cite{todd1995distortions}~\cite{todd2003visual}}. Often only structural properties in 3D left invariant by affine mappings are
reliably perceived, such as the lines/planes parallelism of lines/planes and relative distances in parallel directions. As
a result, 3D glyphs must be designed with great care to convey relationships and patterns, as 2D principles often
do not apply{~\cite{ropinski2011survey}}.}

\begin{figure*}[!tp]
\centering
	\begin{subfigure}[t]{0.33\textwidth}
		\centering
\includegraphics[width=\textwidth]{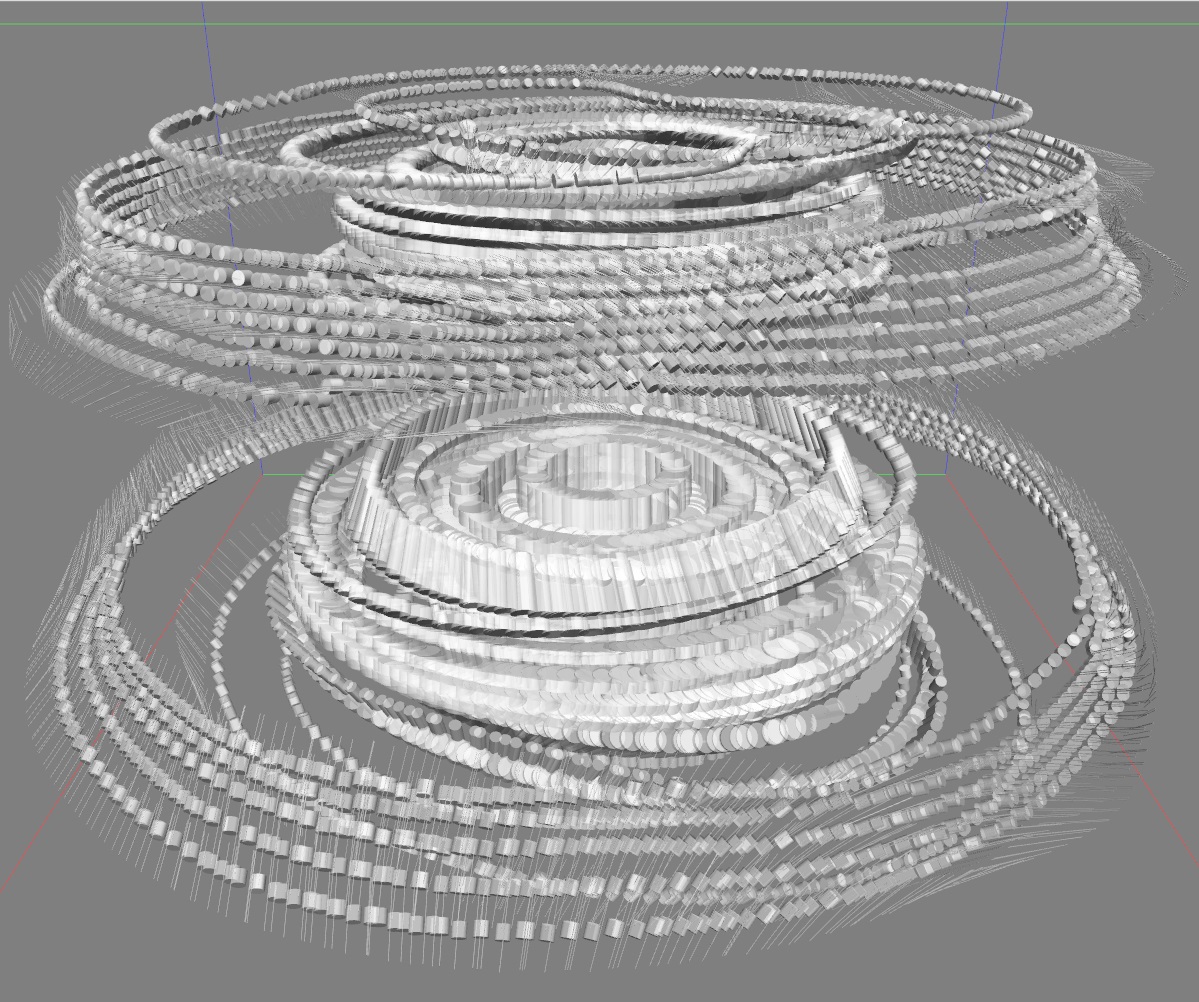}
		\caption{$Length_y$-$length_x$ (L$_y$L$_x$) (integral)}
	\end{subfigure}
	\begin{subfigure}[t]{0.33\textwidth}
	\includegraphics[width=\textwidth]{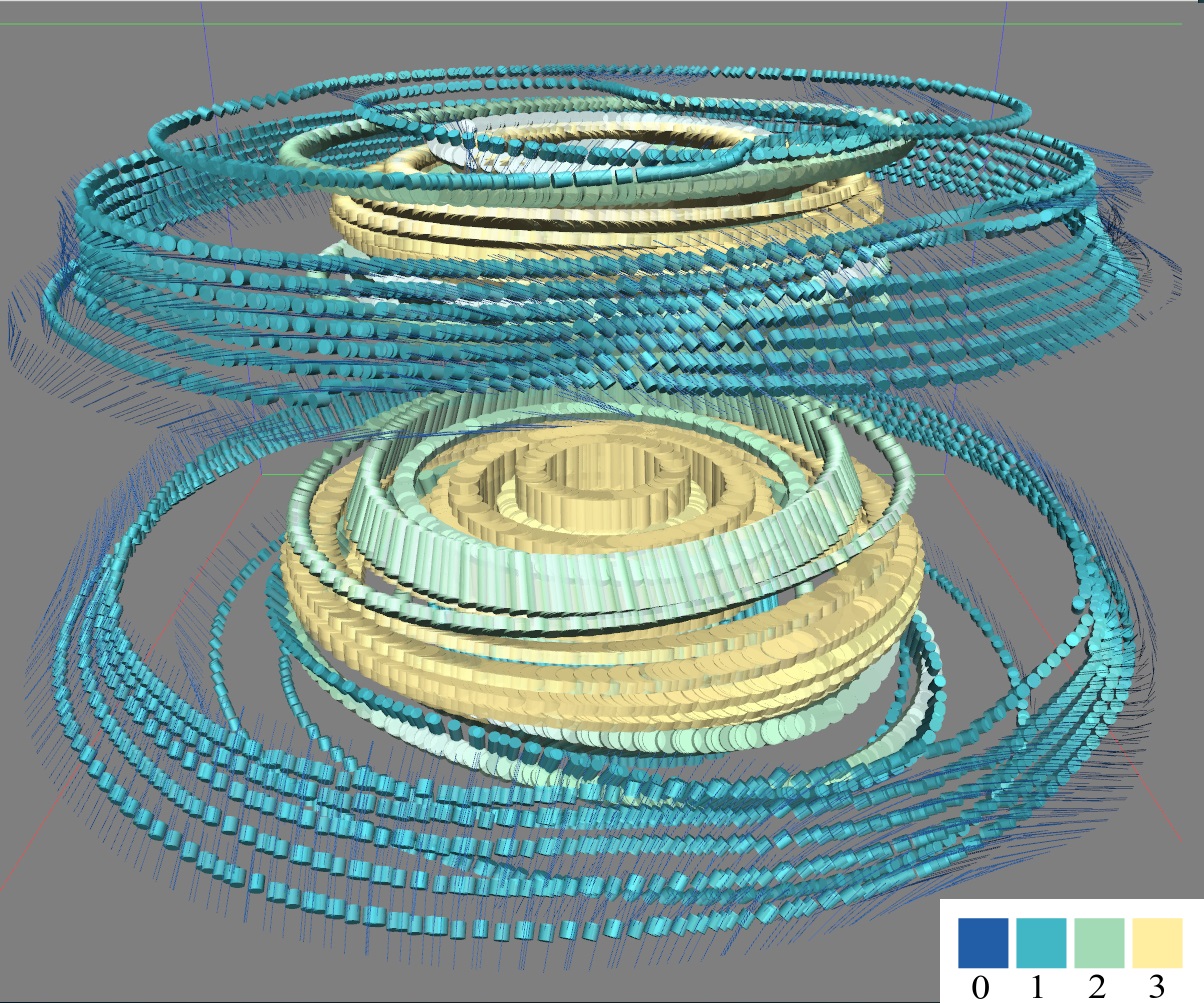}
		\caption{$Length_y$-$color/length_x$ (LCL) (redundant encoding)}
	\end{subfigure}
	\begin{subfigure}[t]{0.33\textwidth}
		\centering
\includegraphics[width=\textwidth]{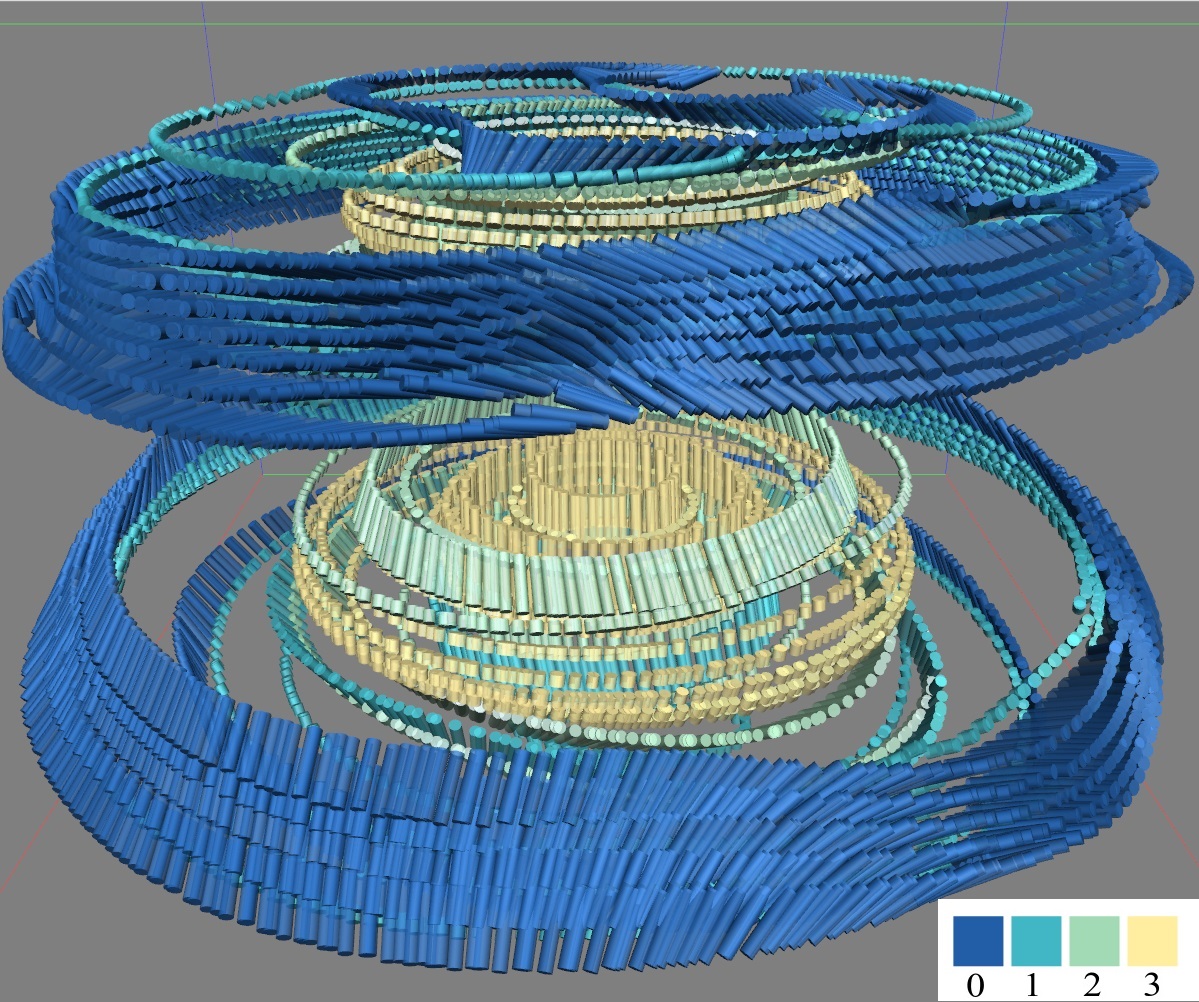}
		\caption{$Length$-$color$ (LC) (separable)}
	\end{subfigure}
    
    	\begin{subfigure}[t]{0.33\textwidth}
		\centering
\includegraphics[width=\textwidth]{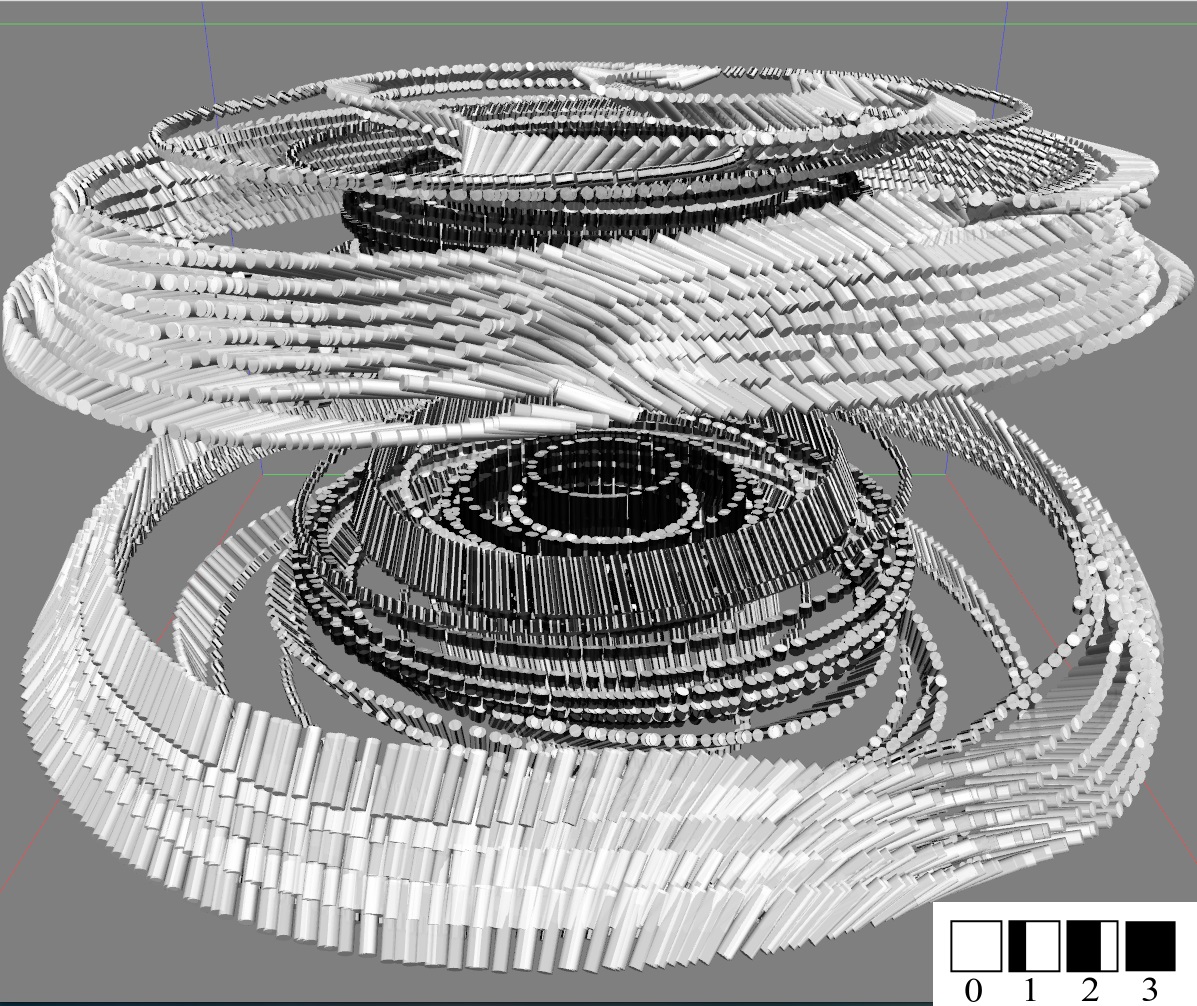}
		\caption{$Length$-$texture$ (LT) (separable)}
	\end{subfigure}
    	\begin{subfigure}[t]{0.33\textwidth}
		\centering
\includegraphics[width=\textwidth]{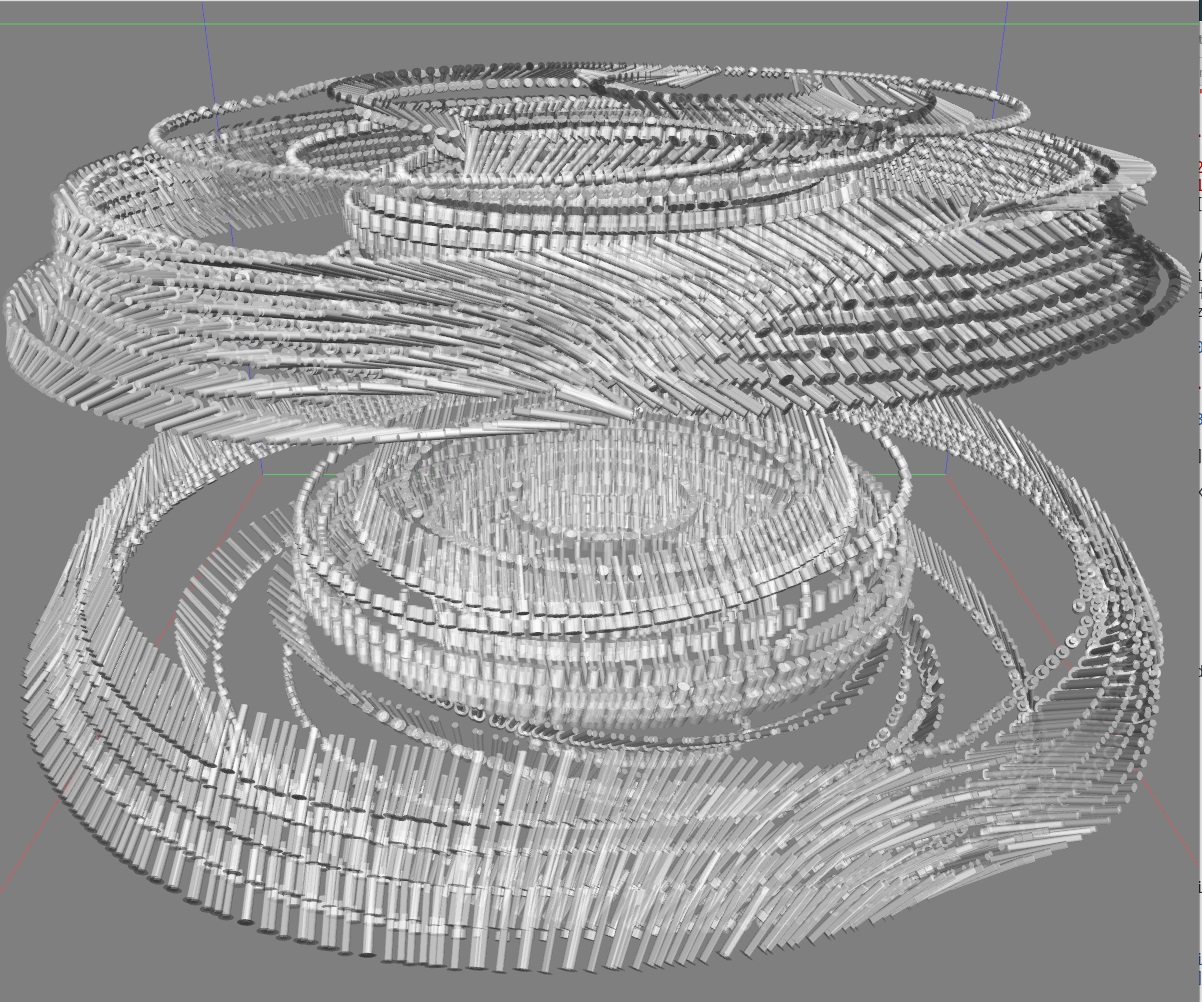}
		\caption{$Length_y$-$Length_y$ (splitVectors) \cite{henan2017}} 
\end{subfigure}
	\begin{subfigure}[t]{0.318\textwidth}    
	\includegraphics[width=\textwidth]{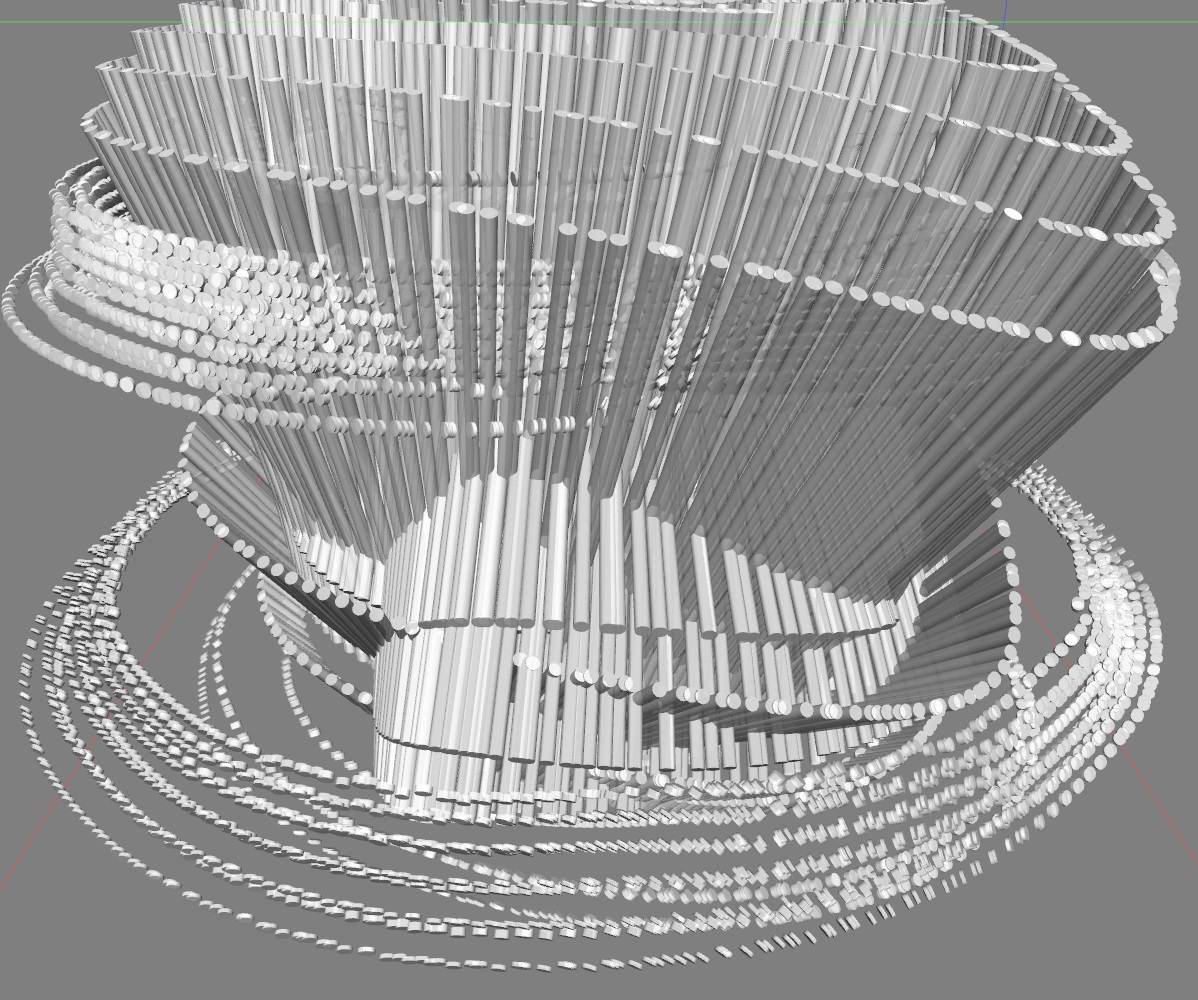}
		\caption{Linear}
	\end{subfigure}
    
\caption{Large-magnitude-range contours computed from a simulation result are shown using five bivariate feature-pairs and linear representation.
}
\label{fig:cases}
\end{figure*}

\remove{Imagine visual search in the 3D large-magnitude-range
vector field, where the differences between the smallest vector magnitude and the largest magnitude reach $10^{12}$.}
On  the  visualization side,  the  initial  design and  evaluation of  large-magnitude-range  vector visualizations use  scientific  notation to  depict  digit  and   power as  two concentric   cylinders{~\cite{henan2017}}:  inside   and   outside tube   lengths ($length_y$-$length_y$) are  mapped to digit  and  power accordingly (aka \textit{splitVectors}, Figure{~\ref{fig:teaser}}e). A three-dimensional (3D) bivariate glyph  scene of this splitVectors design (Figure~\ref{fig:cases}e) achieved up  to  ten times greater  accuracy than  the  traditional direct linear univariate mapping (\textit{linear}) (Figure{~\ref{fig:cases}}f) for reading a vector ratio 
between two  vector  magnitudes. However, this  bivariate splitVectors glyph  also  increases task  completion time  for an apparently simple  comparison task between two vectors in 3D. \textit{Linear} is a significantly more  efficient  approach than their  new solution.

One may frame  this large-magnitude-range issue as a visual design problem: 
\textit{how can we depict a univariate quantity using bivariate visual features or glyphs to help quantum physicists examine complex spatial data?} Intuitively, the last empirical study result  on  vector  magnitude comparisons agrees  well with  a design consensus: to  obtain  a single  magnitude at each  location,   the  human visual   system integrates these  two component parts  (digit  and  exponent terms) into one gestalt.  \add{This integration is referred to as \textit{holistic processing} by Ware{~\cite{ware2012information}} in visualization and  as ~\textit{feature binding} by Treisman{~\cite{treisman1980feature}} in vision science. Both study how our visual system combines separate object  features such  as  shapes, color, motion trajectories, sizes, and distances into the whole object. Ware{~\cite{ware2012information}} (G5.14) further recommends that:  ``\textit{If it is important  for people to respond holistically to a combination of two variables in a set of glyphs, map the variables to integral glyph properties.}''} Since comparison is a \remove{holistic} recognition task,  to represent a univariate vector  magnitude we should always use \textit{integral} properties (visual  properties perceived together as a unit)  or \textit{linear} visualizations, instead of the \textit{separable} (features manipulated and  perceived independently) length pair  in \textit{splitVectors}.

\add{
Essentially, the  current bivariate glyph  design treats  visual  design as a bottom-up stimulus-driven composition in which  the  visual  properties of object  features (e.g.,  orientations and  colors)  are  combined into  single  objects  (here vectors). In  this  work,   we  challenge this consensus and argue that for one,  feature-binding needs not occur
at the  object (vector) level and,  for the other,}  this bivariate splitVectors  gives   viewers  a  \textit{correspondence challenge} that does  not  arise  when integral dimensions or  direct  linear encoding is used - the need to relate
these two quantitative variables to their visual features hampers its efficiency. 
\add{There  are  two  ways to describe human experiences. If the visual field has only one or two objects at a time and if}
splitVectors of length-pairs are used, a viewer  would take longer to process information in order to determine which length is exponent and which is mantissa. 
We suggest that in this case,
if correspondence errors  account for  the  temporal costs with bivariate feature pairs, then techniques preventing this type  of error  can be as effective  as a \change{holistic}{direct} linear  encoding without time-consuming correspondence search.

\remove{
Our method utilizes that fact that binding between separable variables is not always successful and a viewer can thus adopt a sequential task-driven viewing strategy based on visual hierarchy theory{~\cite{wolfe2004attributes}} to obtain gross regional distribution
of larger exponents. 
After this, a lower-order visual comparison within the same exponent can be achieved; And no
binding is needed as long as the correspondence between the
two visual features can be easily understood. With these
two steps, judging large or small or perceiving quantities
accurately from separable variables may be no more time-consuming than single linear glyphs.
}

\add{
Now, for example, if we increase the feature separability by replacing the exponent-to-length mapping in Figures{~\ref{fig:teaser}}e and{~\ref{fig:cases}}e  to exponent-to-color mapping in Figures{~\ref{fig:teaser}}c and{~\ref{fig:cases}}c for comparison tasks, it would be counterproductive for our  attention first  to visit  each  glyph   to  compute the  magnitude (driven by  bottom-up process).  Instead, the  global categorical color (hue)  can guide our  attention to first compare  and  categorize the colors, prior  to a visual  comparison when the colors are the same  (\textit{within} the same  exponent) to compare vector  lengths. In this case, no object-level  binding is needed as long  as the  correspondence between the  two visual  features can be easily  understood. 

Further considering the viewers' task relevant to multiple vector objects (e.g., find  maximum), the same  sequential viewing looking for  the  subregions  followed  by  length-inspection works equally well.  
The  reason is that  feature binding need  not occur  at the object level, but  can be done first  at the scene level, and  scene context benefits  the  reduction of search  regions when there  is  no  ambiguity in  finding  the  correspondences. Coincidentally, this first impression of  the data  to drive statistical information is also  called  
\textit{holistic} or \textit{global 
pattern processing}{~\cite{biederman1977processing}}; Wolfe called features guiding this  \textit{top-down task-driven} attention behaviors as \textit{scene features}{~\cite{wolfe2015guided}}.
Here, we may refer to Ware's holistic features{~\cite{ware2012information}} as \textit{object-level holistic} and Biederman's{~\cite{biederman1977processing}} and Wolfe's{~\cite{wolfe2015guided}}  as \textit{scene-level holistic} design thinking.
}



\remove{
Selecting more separable visual variables to represent a unitary data value
i.e., manipulated and perceived independently,
would initially be considered problematic. 
Often at objective-level, combining two features (here for showing mantissa and exponential terms) into objects (here the magnitudes) is needed.
In our study, choosing separable pairs utilizes the fact that binding between separable variables to form univariate is not always successful, 
a viewer can thus view the salient (exponential) terms,
to obtain gross (regional) distribution;
after this,  visual comparison of mantissa within the same exponent regions can be achieved.
No object-level binding of bivariate features for univariate is needed as long as the global-scene structures formed by the two visual features can be easily understood.
}

\remove{There is a compelling evidence that separable dimensions are not processed together but are broken down into their component parts and processed independently.}
Reducing correspondence error is influenced by the choices of separable dimensions.
According to Treisman{~\cite{treisman1988feature}} and
Wolfe{~\cite{wolfe2018preattentive}}, the initial preattentive phase is the major step towards improved comprehension, more important than the attentive phase. 
\change{Our experiments use}{We select}  the ``most recognizable'' features as color (Figures~\ref{fig:teaser}b, ~\ref{fig:teaser}c, ~\ref{fig:cases}b, ~\ref{fig:cases}c) and texture (Figures~\ref{fig:teaser}d, ~\ref{fig:cases}d), and size (Figures~\ref{fig:teaser}a, ~\ref{fig:cases}a) dimensions. Size and color are preattentive and permit visual selection at a glance, at least in two-dimension (2D). We purposefully select texture patterns by varying
the amount of dark on white, thus introducing luminance variations when many vectors are examined together (Figure{~\ref{fig:cases}}d).
\remove{Our results support that
more separable pairs of $LC$, $LT$, and $LCL$ 
achieve the same temporal cost as the \textit{linear} method.}
Compared to the continuous random noise in Urness et al.{~\cite{urness2003effectively}}, ours is for
discrete quantities and thus 
uses regular scale variations.
When coupled with integral
and separable dimensions, we anticipated that preattentive
pop-out features in separable dimension pairs might reduce
 correspondence errors compared to integral dimensions.
Following this logic,
we hypothesize that \textit{highly distinguishable
separable dimension pairs might erase the costs associated with the correspondence errors to reduce task completion time and be more accurate.} 

We tested this hypothesis in \change{an}{two} experiments with  
\change{four tasks}{six tasks} using 
four dimension pairs to compare against
the $length_y$-$length_y$ (separable) in Zhao et al.~\cite{henan2017}: 
$length_y$-$length_x$ (integral), $length$-$color$ (separable), $length$-$texture$ (separable), and $length_y$-$color/length_x$ (redundant and separable).
Since we predicte that separable dimensions with more preattentive features would reduce the task completion time, 
$length$-$color$ and $length_y$-$color/length_x$
might achieve more efficiency without hampering accuracy than other bivariate feature-pairs.

This work makes the following contributions: 

\begin{itemize}

\item Empirically validates that bivariate-glyphs encoded by highly \remove{distinguishable} separable dimensions would \change{reduce correspondence errors}{improve comparison task complete time} (Exp 1). 


\item 
\add{Is the first to explain the benefits of the \textit{global scene-guidance} which
expands the widely accepted 
\textit{object-level} bivariate glyph design in visualization (Exp 2).
}

\item
Offers a rank order of separable variables for 3D glyph design and 
shows that the separable pairs $length$-$color$ and $length$-$texture$ are among the most effective and efficient \change{glyph encodings.}{feature pairs.}

\end{itemize}

\section{Theoretical Foundations in Perception and Vision Sciences}

At least three perceptual and vision science theories have inspired our work: \textit{integral and separable dimensions}~\cite{garner1970integrality}.
preattentive features ranking~\cite{healey1999large, healey1995visualizing, mackinlay1986automating, cleveland1984graphical}, and \textit{monotonicity}~\cite{ware2009quantitative}. 

\remove{\textbf{Terminology.}
To avoid confusion, we adapt terms from Treisman and Gelade{~\cite{treisman1980feature}} in vision science
to visualization. We use ``visual dimension'' to refer to the complete range of variation that is separately 
analyzed by some functionally independent perceptual subsystem, and ``visual feature'' to refer to  a particular  value  
on  a  dimension.  Thus  color, texture, and size  are visual dimensions; gray-scale, spatial-frequency, and length are the features on those dimensions. 
Our ``visual dimension'' is thus most similar to Bertin's ``visual variables''{~\cite{bertin1967semiology}} }
\remove{in the visualization domain. 
Differentiating the terms \textit{dimension} and \textit{features} is necessary for us in the long-term to compare design both within (as features) and between the dimensions.}

\textbf{Integral and Separable Dimensions.}
Garner and  Felfoldy's  seminal work on integral and separable dimensions~\cite{garner1970integrality} has inspired many visualization design guidelines. 
Ware~\cite{ware2012information} suggests a continuum from more integral to more separable pairs: 
\textit{(red-green)}-\textit{(yellow-blue)}, \textit{$size_x$}-\textit{$size_y$},  \textit{color-shape/size/orientation}, \textit{motion-shape/size/orientation}, 
\textit{motion-color}, and \textit{group position-color}. 
His subsequent award-winning bivariate study~\cite{ware2009quantitative} using \textit{hue-size}, \textit{hue-luminance}, and \textit{hue-texton} (texture) 
supports the idea that more separable dimensions of \textit{hue-texton} lead to higher accuracy. 
Our work follows the same \change{choices}{ideas of applying integral and separable dimensions} but differs from Ware's texton selection in 
two important aspects. 
\remove{the dependencies between two variables to be represented and whether or not the texture encodes continuous and ordered data.}
First,
the Ware study focuses on finding relationships between two \textit{independent} data variables, and thus his tasks are analytical;
\remove{In contrast, ours demands two component parts from  
a unitary variable represented in two parts.}
Second, our texture uses the amount of black and white to show \change{continuous local spatial frequency,}
{luminance variations,} in contrast to the discrete shape variation in textons.
We anticipate that ours will be more suitable to continuous
quantitative values{~\cite{wolfe2004attributes}}.
No existing work we know of has studied whether or not the separable features can facilitate global comparisons and can be scaled to 3D vector field analysis.

\textbf{Feature-Binding and Scene-Guidance Theories.}
Treisman and Gelade’s feature-integration theory of attention~\cite{treisman1988feature} showed that the extent of difference between
target and distractors for a given feature affects search time.
This theory may explain why splitVectors was time
consuming: the similarity of the two lengths may make them
interfere with each other in the comparison, thus introducing temporal cost. What we ``see'' depends on our goals
and expectations. Wolfe et al. propose the theory of ``\textit{guided
search}''~\cite{wolfe2015guided, wolfe2007guided}, a first attempt to incorporate users' goals
into viewing, suggesting that what users see is based on users' goals. Wolfe et al. further suggest that
color, texture, size, and spatial frequency are among the most
effective features in attracting the user's attention.

\remove{Building on these research, our current study shows
that viewers can be task-driven and adopt optimal viewing
strategies to be more efficient. No existing visualization
work to our knowledge has studied how viewers' strategies
in visual search influence bivariate visualization of two
dependent variables. While Ware has recommended holistic
representations for holistic attributes, our empirical study
results suggest the opposite: that separable pairs can be as
efficient as holistic representations.
}

\textbf{Preattentive and Attentive Feature Ranking.}
Human visual processing can be faster when it is preattentive, i.e., perceived before it is given focused attention~\cite{treisman1988feature}. 
The idea of pop-out highlighting of an object is compelling because it captures the user's attention against a background of other objects
(e.g., in showing spatial  highlights~\cite{strobelt2016guidelines}).
Visual features such as orientation and color (hue, saturation, lightness) can generate pop-out effects~\cite{treisman1988feature}~\cite{healey1996high}.
Healey and Enns~\cite{healey2012attention} 
in their comprehensive review further
remark that
these visual features are also not popped-out at the same speed:
\textit{hue} has higher priority than \textit{shape} and \textit{texture}~\cite{callaghan1989interference}.

Visual features also can be responsible for different attention speeds, and color (hue) and size (length and spatial frequency) are among those that guide attention{~\cite{wolfe2004attributes}}.
For visualizing quantitative data, MacKinlay~\cite{mackinlay1986automating}    and Cleveland and McGill~\cite{cleveland1984graphical} leverage the 
ranking of visual dimensions and suggest that position and size are quantitative and can be compared \add{in 2D}. 
Casner~\cite{casner1991task}
expends  MacKinlay's APT by incorporating user tasks to guide visualization generation. 
Demiralp et al.~\cite{demiralp2014learning} evaluate a crowdsourcing
method to study subjective perceptual distances of 2D bivariate pairs of shape-color, shape-size, and size-color. 
When adopted in 3D glyph design, these studies further suggest that the most important data 
attributes should be displayed with the most salient 
visual features, to avoid situations in which secondary data values mask the information the viewer wants to see.

\textbf{Monotonicity.}
Quantitative data encoding must normally 
be monotonic, and various researchers have recommended a coloring sequence that increases monotonically in luminance{~\cite{rogowitz2001blair}}. 
In addition, the visual system mostly uses luminance variation to determine shape information{~\cite{o2010influence}}. There has been much debate about the proper design of a color sequence for displaying quantitative data, mostly in 2D{~\cite{harrower2003colorbrewer}}  and in
3D shape volume variations{~\cite{zhang2016glyph}}.
Our primary requirement is that users be able to read large or
small exponents at a glance. We chose four color steps \add{in the first study and up to seven steps in the second study} for showing areas of large and small exponents that are mapped to a hue-varying sequence. 
\remove{monotonic luminance and the higher the luminance, the higher exponents.} We claim not that \change{this color sequence}{these color sequences} are optimal, only that they are reasonable solutions to the design problem{~\cite{harrower2003colorbrewer}}. 



\section{Experiment I: Local Discrimination and Comparisons}

\add{The goal  in this  first experiment is to quantify the  benefits of separable pairs  for visual  processing of a few items.  This section discusses the experiment, the design knowledge we can gain  from it, and  the factors  that  influence our design.
}

\subsection{Methods}
\subsubsection{Bivariate Feature-Pairs}

We choose five bivariate feature-pairs 
to examine the comparison task efficiency of separable-integral pairs.

$Length_y$-$length_x$ (\textbf{\textit{integral}}) (Figure~\ref{fig:teaser}a). 
Lengths encode digits and exponents shown as the diagonal and height of the cylinder glyphs.

$Length_y$-$color/length_x$ (\textbf{\textit{redundant and separable}}) (Figure~\ref{fig:teaser}b). This pair compared to $length_y$-$length_x$ adds
a redundant color (luminance and hue variations) dimension to the exponent and  the four sequential colors are chosen from
Colorbrewer~\cite{harrower2003colorbrewer}. 

$Length$-$color$ (\textbf{\textit{separable}}) (Figure~\ref{fig:teaser}c). This pair 
maps \remove{four} exponents to color. Pilot testing shows that correspondence errors in this case would be the lowest among these five feature-pairs.

$Length$-$texture$ (\textbf{\textit{separable}}) (Figure~\ref{fig:teaser}d). Texture represents exponents. The percentage
of black color (Bertin{~\cite{bertin1967semiology}}) is used to represent the exponential
terms 0 ($0\%$), 1 ($30\%$), 2 ($60\%$) and 3 ($90\%$), wrapped around the cylinders in  five segments to make them visible from any viewpoint.

$Length_y$-$length_y$ (\textbf{\textit{splitVectors~\cite{henan2017}}}, \textbf{\textit{separable})} (Figure~\ref{fig:teaser}e). This glyph uses  splitVectors~\cite{henan2017} as the baseline and maps both digit and exponent to lengths. The glyphs are 
semitransparent so that the inner cylinders showing the digit terms are legible.

\textit{Feather-like fishbone legends} are added at each location when the visual variable
\textit{length} is used. The \textit{tick-mark band} is depicted as subtle light-gray lines around each cylinder. 
Distances between neighboring lines show a unit length legible at certain distance (Figure~\ref{fig:teaser}, rows 2 and 3).

\subsubsection{Hypotheses}

Given the analysis above and recommendations in the literature, we arrived at the following working hypotheses:

\begin{itemize}

\item
\textit{Exp I.H1. (Overall). The $length$-$color$ feature-pair can lead to the most accurate answers.}

Several reasons lead to this conjecture. Color and
length are separable dimensions. Colors can be detected quickly, so length and color are highly distinguishable. Compared to the redundant $length_y$-
$color/length_x$, $length$-$color$ reduces density since the \change{glyphs}{feature-pairs} are generally smaller than those in $length_y$-$color/length_x$.

\item 
\textit{Exp I.H2. (Integral-separable).
Among the three separable dimensions, $length$-$color$ may lead to the greatest
speed and accuracy and $length$-$texture$ would be more effective than splitVectors.}

The hypothesis could be supported because color
and length are highly separable.


\item \textit{Exp I.H3. (Redundant hypothesis).
The redundant pair $length_y$-$color/length_x$ will reduce time compared to splitVectors}.

\add{This hypothesis could be supported because 
redundancy increases information processing capacity.}

\end{itemize}


\subsubsection{Tasks}

Participants perform the following three task types \add{as in Zhao et al.{~\cite{henan2017}} so that results are comparable.} They had unlimited time to perform these three tasks.

\begin{figure}[!tb]
	\begin{subfigure}[t]{\columnwidth}
 \centering 
 \includegraphics[width=\columnwidth]{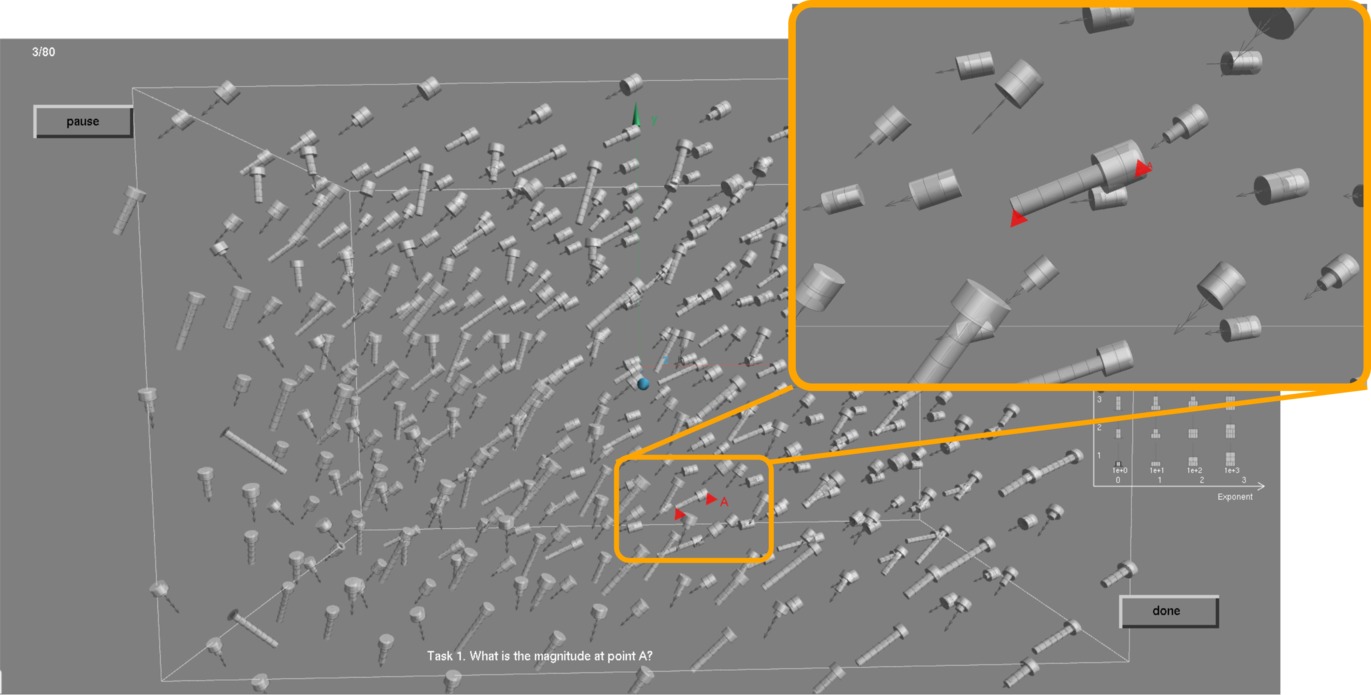}
 \caption{Exp1 MAG task: What is the magnitude of the vector at point A? (answer: 636.30)}
 \label{fig:task1}
    \end{subfigure}

	\begin{subfigure}[t]{\columnwidth}
 \centering 
 \includegraphics[width=\columnwidth]{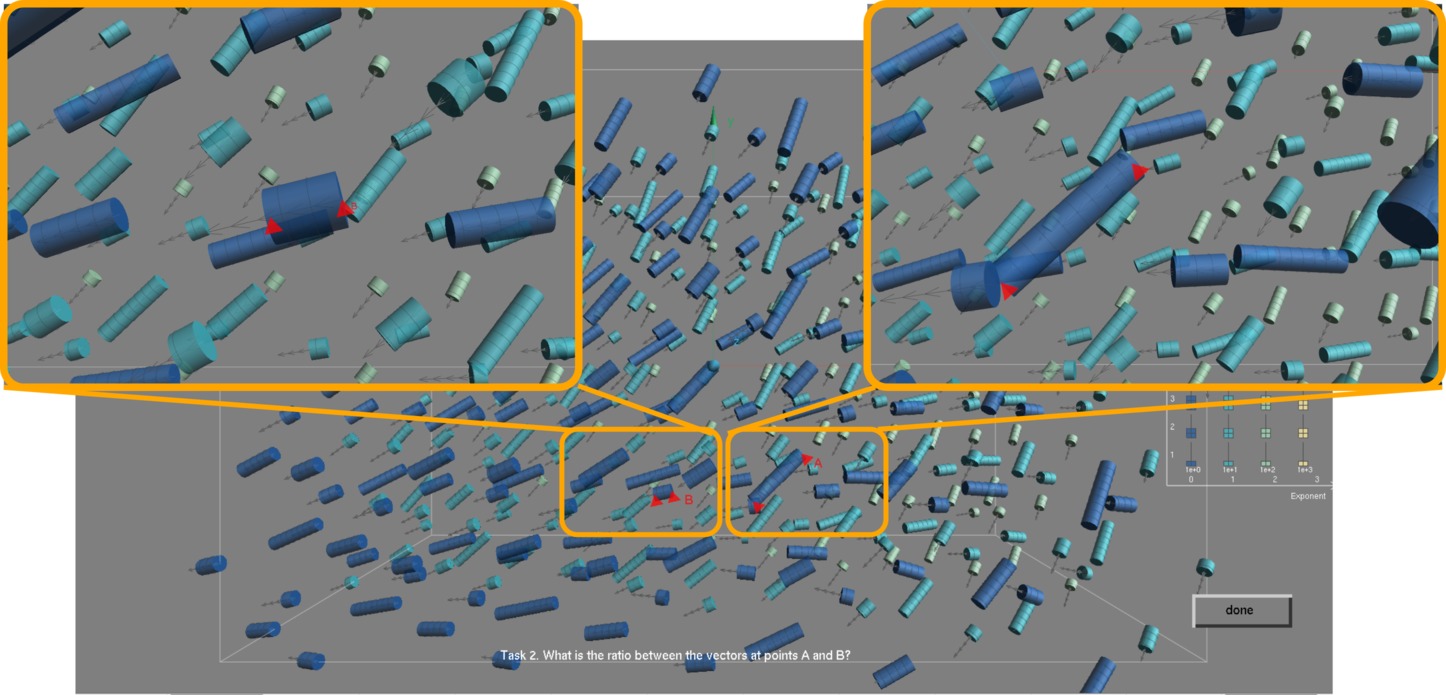}
 \caption{Exp1 RATIO task: What is the ratio of the magnitude between the vectors at points A and B? (answer: 3.60)}
 \label{fig:task2}
\end{subfigure}

	\begin{subfigure}[t]{\columnwidth}
 \centering 
 \includegraphics[width=\columnwidth]{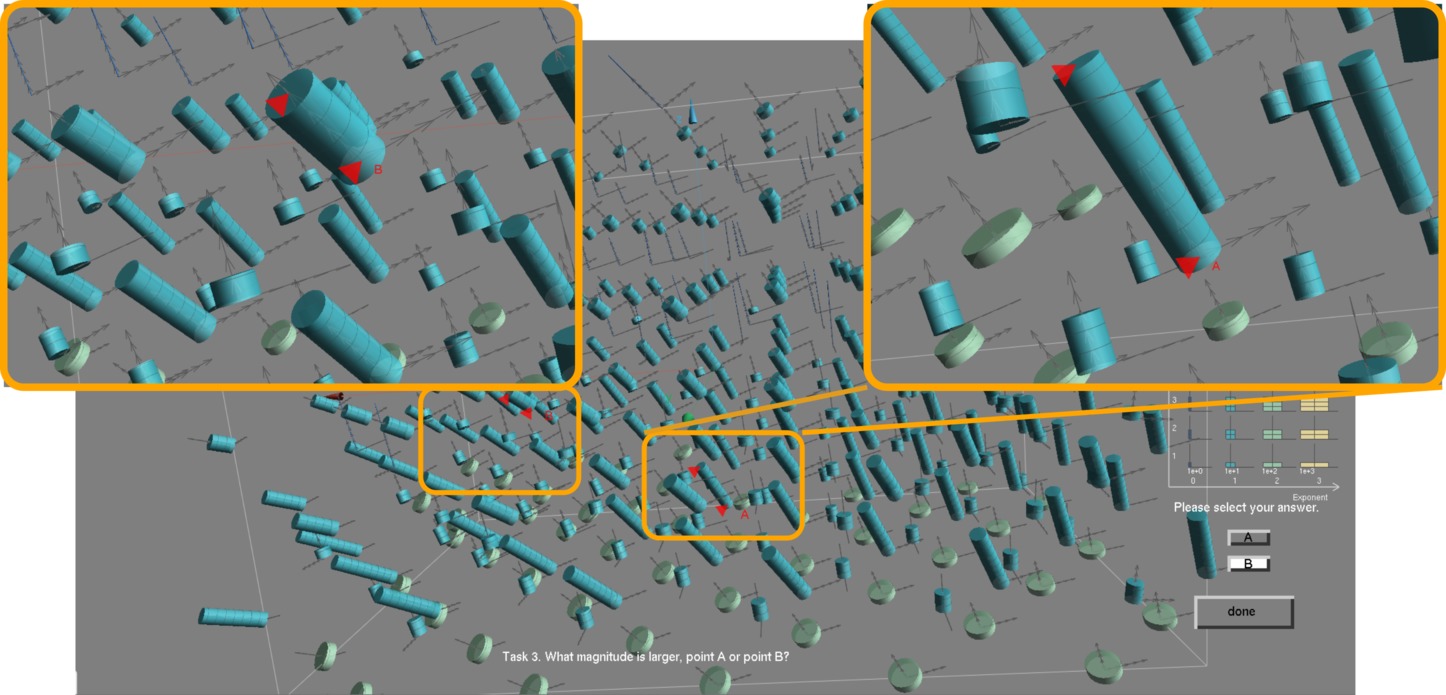}
 \caption{Exp1 COMP task: Which magnitude is larger,  point A or point B? (answer: A on the right.)}
 \label{fig:task3}
\end{subfigure}

\caption{Experiment 1: Local discrimination and comparison tasks.}
\label{fig:expTasks}
\end{figure}

\textbf{Exp1.Task 1 (MAG):  magnitude reading (Figure~\ref{fig:task1})}. \textit{What is the magnitude  at  point  A?} 
One  vector  is  marked  by  a  red  triangle labeled ``A'', and participants should report the magnitude of that vector. This task requires precise
numerical input.

\textbf{Exp1.Task 2 (RATIO):   ratio  estimation (Figure~\ref{fig:task2})}. \textit{What is the ratio of magnitudes of points A and B?}
Two vectors are marked with two red triangles labeled ``A'' and ``B'', and participants should estimate  the  ratio  of  magnitudes  of  these  two
vectors. 
The ratio judgment is the most challenging quantitative task~\cite{mackinlay1986automating}. Participants can either compare the glyph shapes or decipher each vector magnitude and compute the ratio mentally.

\textbf{Exp1.Task 3 (COMP):  comparison (Figure~\ref{fig:task3})}. \textit{Which magnitude is larger,  point A or B?} 
Two vectors are marked with red triangles and labeled ``A'' and ``B''. Participants select their answer by directly clicking the ``A'' or ``B'' answer buttons. This task is a simple comparison between two values and offers a binary choice of large or small.

\remove{
\textbf{Exp1.Task 4 (MAX):  identifying  the  extreme  value within  30 seconds (Figure{~\ref{fig:task4}})}.
\textit{Which point has maximum magnitude when the exponent is X?}
X in the study was a number from 0 to 3. Participants need first to locate points with exponent X and then select the largest one of that group.
Compared to Task 3, this is a global task requiring participants to find the extreme among many vectors.
}

\subsubsection{Data Selection}
\label{sec:dataselection}
Because we are interested in comparing our results to those in Zhao et al.{~\cite{henan2017}}: We \add{replicate their data selection method} to generate the data by randomly sampling some quantum physics simulation results and produce samples within 3D boxes of size $5\times3\times3$. 
There are 445 to 455 sampling locations in each selected data region.

We select the data satisfying the same following conditions: (1) the answers must be at locations where some context information is available, i.e., not too close to the boundary of the testing data. (2) no data sample is repeated to the same participant; 
(3) Since data must include a broad measurement, we select the task-relevant data from 
each exponential term of 0 to 3.

\remove{
For task 1 (MAG, \textit{What is the magnitude at point A?}), point A was in the range $[-1/3, 1/3]$ of the center of the bounding box in each data sample. 
In addition, the experiment had four trials for each variable pair
with one instance of the exponent values of 0, 1, 2 or 3 being used.
}
\remove{
For task 2 (RATIO, \textit{What is the ratio of the magnitudes of points A and B?}) points A and B are again randomly selected; the choice of exponents
is the same as task 1 as well. Thus the ratios were always larger than 1. 
}
\remove{
For task 3 (COMP, \textit{Which magnitude is larger, point A or point B?}), points are again must be in the range $[-1/3, 1/3]$ of the center of the bounding box.
The magnitude of one point is around $Max_{magnitude}$ $\times$ 0.2, and magnitude of the other point is around $Max_{magnitude}$ $\times$ 0.5 where $Max_{magnitude}$ is the maximum magnitude in the data sample used for the corresponding trial.
}
\remove{
For task 4 (MAX, \textit{Which point has maximum magnitude when the exponent is X?}) X was an instance of the exponent values in the exponent range. In the first study, the range was fixed to 4.
}

\subsubsection{Empirical Study Design}


\textbf{Design and Order of Trials.} We use a within-subject design with one independent variable of bivariate quantitative \change{glyphs}{feature-pair} (five types).\remove{and compared their efficiency in four tasks.}
Dependent variables are relevant error \add{(for MAG and RATIO)} or accuracy \add{(for COMP)} and task completion time. We also collect participants' confidence levels.
\add{
The
\textit{accuracy measure} follows Zhao et al.{~\cite{henan2017}} to study how sensitive a method is to error uncertainty based on the relative error (RE) or fractional uncertainty, calculated as \textit{RE = $\lvert$ correct answer - participant answer $\rvert$ / (correct answer)}. This measure is used for MAG and RATIO tasks. The benefit of this approach is that it takes into account the value of the quantity being compared and thus provides an accurate view of the errors. 
}
\input{exp1design.tex}

Table~\ref{tab:experimentdesign} shows that participants are assigned into five blocks in
a Latin-square order, and within one block the order of the five \change{glyph}{feature-pair} types is the same.
Participants perform tasks with randomly selected datasets.\remove{for each encoding on each task type.}
Each participant performed 
$60$ \change{subtasks}{trials} ($3$ tasks $\times$ $4$ random data $\times$ $5$ \change{bivariate-glyphs}{feature-pairs}). \remove{We ran four trials for each encoding method exponent.}
\add{These four random data are from four exponent ranges.}

\textbf{Participants.}
We diversify the participant pool as much as possible, since all tasks can be carried out by those with only some science background. 
Twenty participants (15 male and 5 female, 
mean age = 23.3, and standard deviation = 4.02) participated in the
study, with ten in computer science, three in engineering,
two in chemistry, one in physics, one in linguistics, one
in business administration, one double-major in computer
science and math, and one double-major in biology and
psychology. The five females are placed in each of the five
blocks (Table~\ref{tab:experimentdesign}). 
On average, participants spent about 40 minutes on the computer-based tasks.

\textbf{Procedure, Environment, and Interaction.}
Participants  are  greeted  and  complete  an  Institutional  Review Board (IRB) consent form. All participants had 
normal or corrected-to-normal vision and passed the Ishihara color-blindness test. They filled in the informed consent form (which described the procedure, risks and benefits of the study) and the demographic survey.
We showed \change{glyph}{feature-pair} examples and trained the participants with one trial for \change{each of the five glyphs}{every feature-pair} per task. \add{They were told to be as accurate and as quickly as possible, and that accuracy was more important than time.} They could ask questions during the training but were told they could not do so during the formal study. Participants practiced until they fully understood the \change{glyphs}{feature-pairs} and tasks. After the formal study, participants filled in a post-questionnaire asking how these feature pairs supported their tasks and were interviewed for their comments.

Participants  sat  at  a $27\,''$ BenQ  GTG  XL  2720Z,
gamma-corrected display with resolution 1920 $\times$ 1080. The distance  between  the  participants  and  the  display  was  about
$50cm$. The minimum visual angle of task-associated glyphs was $0.2^{\circ}$ in the default view where all data points were visible and filled the screen. 

\add{Participants could rotate the data and zoom in and out. Lighting placement and intensity were chosen to produce visualization with contrast and lighting properties appropriate for human assumptions and the spatial data. The screen background color was neutral stimulus-free gray background to minimize the discriminability and appearance of colors~\cite{ware2012information}. Using black or white background colors will make the black and white texture stimuli disappear thus bias the results.}

\subsection{Experiment I: Results and Discussion}

\input{Exp1timeError.tex}

\subsubsection{Analysis Approaches} 

\change{
We collected 1600 data points (80 from each of the 20
participants), and there were 400 data points from each of
the four tasks.} {We collected 400 data points for each task.}
\add{In preparing the accuracy and task completion time for analysis, a trial was considered to have an answer of the \textit{first type of correspondence error}  if responses' \textit{exponent} value did not match the correct one for the MAG task. 
This correspondence errors occurred when participants had trouble differentiating the levels within a encoding.}

\add{We detected 11 instances of the \textit{first type of correspondence errors} from MAG (these trials comprised $2.75\%$ of the total: three splitVectors, five $length_y$-$length_x$, one $length$-$texture$, and two $length_y$-$color/length_x$). 
This correspondence error appeared to be influenced by the integral-separable dimension as well and the integral dimension $length_y$-$length_x$ had the highest (5) and $length$-$color$ had no instances. 
We used only the remaining \textit{correct} ones in the statistical analysis because these errors would mask all other data by being at least one order of magnitude larger.
For the remaining data in MAG and all data in RATIO and COMP tasks, we used standard outlier detection by first calculating the mean and standard derivation across all trials for each participant and pruning any trials that were +/- two standard derivations from that participant's mean. With this approach, no outlier was detected in the MAG, RATIO, and COMP tasks.}

Table{~\ref{tab:new_glm}} and Figure{~\ref{fig:timeError}} show the $F$ and $p$ values computed with SAS one-way measures of variance for task completion time. \remove{($log_{e}(time)$) base to obtain a normal distribution), the Friedman test of accuracy, and repeated measures of logistic regression on confidence levels. Post-hoc analyses on $log_{e}(time)$ are adjusted by Bonferroni correction.} 
\add{A post-hoc analysis using Tukey's Studentized Range test (HSD) was performed when we observed a significant main effect. 
When the dependent variable was binary (i.e., answer correct or wrong), we used a logistic regression and reported the \textit{p} value from the Wald $\chi^2$ test.
When the $p$ value was less than 0.05, variable levels with $95\%$ confidence interval of odds ratios not overlapping were considered significantly different.
}
All error bars represent $95\%$ confidence intervals.
We also evaluated effect sizes \change{using Cohen's $d$ for continuous data (e.g., time), and Cramer's V for binary choice (e.g. accuracy) to understand practical 
significance{~\cite{cohen1988statistical}}. 
We used Cohen's benchmarks for ``small'' ($0.07-0.21$), ``medium'' ($0.21-0.35$), and ``large'' ($>0.35$) effects.}
{using \textit{eta-square}, labeled ``small'' $(0.01-0.06)$, ``medium'' $[0.06-0.14)$, and ``large'' $\geq 0.14$ effects following Cohen~\cite{cohen1988statistical}.
}


\subsubsection{Overview of Study Results}
\remove{All hypotheses but H2 are supported.}
Our results clearly demonstrated the benefits  \add{in terms of task completion time} of separable
dimensions for comparison. 
We observed a significant main effect of feature-pair type on task completion time for all three tasks MAG, RATIO, and COMP, and the effect sizes were in the medium range (Table~\ref{tab:new_glm}, Figure~\ref{fig:timeError}). 
$Length$-$color$ was the most efficient approach.\remove{ and had the least error.} For \change{the comparison tasks (COMP and MAX) in this study}{COMP}, $length$-$color$, $length$-$texture$ and $length_y$-$color/length_x$ were most efficient for simple two-point comparison (Figure{~\ref{fig:task3timeError}}).\remove{ and were most accurate for 
group comparisons (Figure{~\ref{fig:task4timeError}}).}
\remove{
 A most surprising result was that both $LC$ and $LT$ were highly accurate and efficient.}

\subsubsection{Separable Dimensions Are Better Than Integral Dimensions for Local Comparisons}

Our separable-integral hypothesis (H1) was supported.
In the MAG tasks, the \textit{integral} $length_y$-$length_x$ was least efficient and all other separable-pairs were in a separate group, the most efficient one (Figure~\ref{fig:task1timeError}). 
In the RATIO tasks, $length$-$color$, $length$-$texture$, and $splitVectors$ were the most efficient group (Figure~\ref{fig:task2timeError}); in the COMP tasks, the \textit{redundant} $length_y$-$color/length_x$,  $length$-$color$, and $length$-$texture$ were in the most efficient group (Figure~\ref{fig:task3timeError}). 

\textit{SplitVectors} was not as bad as we originally thought in handling correspondence errors, especially for the quantitative reading
tasks of MAG and RATIO. \textit{SplitVectors} belonged to the same efficient post-hoc group as $length$-$color$ and $length$-$texture$ for the
RATIO tasks and these three were also most efficient for MAG.
\remove{
The MAG and RATIO are the only two quantitative tasks. In contrast,
The $length_y$-$length_x$ pairs did elongate the task completion time.
}

\add{We speculate that  this result  may indicate that  when the comparison set  size  was  small,  participants did  not  need scene-level information to achieve  accuracy. We anticipate that  when the  search  space  set-size  increases, the  search will become  time-consuming and the  lack  of  scene-level features would increase correspondence error  and  thus  reduce  effectiveness. We observe this in Experiment II.}

\remove{
The general order of these three separable visual variable pairs was that more separable pairs were more efficient; 
And the efficiency and effectiveness of these feature-pairs were very much task dependent. One of the most interesting results is that Separable $length$-$color$ and $length$-$texture$ resulted in high efficiency in nearly all tasks:
$length$-$texture$ functioned just as well as the $length$-$color$ with comparable 
subjective confidence levels. 
This result can be explained that the black/white texture scales on a regular grid may lead to luminance variation, which
attracts attention{~\cite{wolfe2004attributes}}
thus directly contribute to discrimination of the \textit{global and spatial} pattern differences.}

\subsubsection{Separable pairs of $length$-$color$ and $length_y$-$color/length_x$ achieved comparable efficiency to direct linear glyph}

Critical for motivating this experiment was whether the separable pairs supported COMP and how the separable pairs compared in efficiency to the direct mapping. 
Since our study had the same numbers of sample data as Zhao et al.~\cite{henan2017},
we then performed a one-way $t$-test 
to compare against the
direct linear encoding in Zhao et al.~\cite{henan2017}. 
Our separable-hypothesis (H2) was supported and our results indicated that COMP
(judging large or small) from separable variables
was no more time-consuming than direct linear glyphs.
Our post-hoc analysis 
showed that $length$-$color$, $length$-$color/length$, and $linear$ were in the same post-hoc group, i.e. that there were no significant differences between these features.
We also observed that \textit{splitVectors} dropped to the least efficient or most error-prone post-hoc groups  (Fig.~\ref{fig:task3timeError}). This result replicated the former study results in Zhao et al.~\cite{henan2017} 
by showing that splitVectors impaired comparison efficiency.
\remove{or effectiveness.}

This result may be explained by the idea that the highly separable
pairs may turn the comparison into a single-dimension
digit comparison tasks, since a viewer could quickly resolve the two exponents and thus reduce the correspondence error introduced by the \textit{splitVectors} design.

 



\remove{
\textbf{Relative Error or Accuracy.}
We adopted the error metric for quantitative data of  
Cleveland and McGill{~\cite{cleveland1984graphical}} for task types MAG, RATIO, and MAX.
This metric calculates the absolute difference between the user's 
and the true difference using 
the formula
$log_{2} {(| judged\,percent - true\,percent | + \frac{1}{8} ) }$, where the $log_2$
base was appropriate for relative error judgments and $\frac{1}{8}$ prevented distortion of the results towards
the lower end of the error scale, since some of the absolute errors were close to 0. 
}


\remove{
Participants ranked their confidence levels after each trial during the computer-based study. Preferences were collected in the post-questionnaire. Both data were on a scale of 1 (least confident or preferred) to 7 (most confident or preferred). 
Significant effects of the glyph type on confidence were only
observed in the holistic comparison task of MAX, but not in local tasks of MAG, RATIO, and COMP. 
$LC$ was the top preferred glyph for all tasks followed by $LCL$ and then $LT$. The two length-based regardless orthogonal or parallel were least preferred. The confidence levels followed a similar trend as the preferences.
} 

\subsubsection{Redundant Feature-Pairs Were Efficient}

\remove{We might also compare the coloring effect with that of Healey, Booth, and Enns{~\cite{healey1996high}}.  
Their single-variate study showed that color was strongly influenced by the surroundings of 
the stimulus glyph, caused a significant interference effect when participants had to judge heights of glyphs or 
density patterns. We did not observe such effects here because 
the colors are discrete and can be easily distinguished.
}

\remove{
We tested this conjecture through our observations with some new quantum physics simulation datasets from our collaborators as shown in Figure{~\ref{fig:cases}}. We can easily discriminate
the boundaries between the adjacent magnitude variations in 
the $length_y$-$texture$ (Figure{~\ref{fig:cases}}e) and $length_y$-$color$ (Figure{~\ref{fig:cases}}d) 
feature boundaries and these two share a similar effect.
}

\remove{We think that $length_y$-$length_y$ was an effective and efficient feature-pair for quantitative tasks because
the same type used in the glyph perhaps reduced the cognitive load and also because 
scales of parallel lines are preserved in 3D.
}


\remove{It is worth noting that the only difference among 
these four tasks was that the first two (MAG and RATIO) involve visual discrimination (knowing precise values or how 
much larger) and COMP involved visual detection (larger or higher). For MAG and RATIO, a long time may have been 
spent on mentally synthesizing the numerical values at individual locations. 
Our results further confirmed that \textbf{\textit{visual discrimination} and \textit{visual detection} were 
fundamentally different comparison tasks} as shown in Borgo et al.{~\cite{borgo2014order}}.
}

\remove{
\textbf{The errors or accuracy was task-dependent and perhaps depends on set-size.}
The lack of significant main effect on errors or accuracy happened in all tasks (MAG. RATIO, and COMP). 
Note that none of these three tasks required initial visual search, and target
answers were labeled. Wolfe called this type of task-driven with known-target \textit{guided} tasks{~\cite{wolfe2007guided}}.
$Length_y$-$color$ was most accurate in all task types. We thought at first that error may be related to 
so-called \textit{proximity}, i.e., the perceptual closeness of 
visual variable choices to the tasks. The coloring was perhaps more direct. However, since the 
participants read those quantities as they commented, we thought the reason for not observing difference could well be
their similarities in mentally computing cost.
\textbf{When search-space set-size increases for the MAX tasks, the search becomes time-consuming and 
none of the length pairs ($length_y$-$length_y$ and $length_y$-$length_x$) was effective.}
}

We also confirmed hypothesis H3. We were surprised by the large performance gain with the redundant 
encoding $length_y$-$color/length_x$ of mapping $color$ and $length$ to the exponents in splitVectors. 
With redundant encoding, 
the relative error was significantly reduced and task completion time was much shorter
(significantly shorter for MAG and COMP tasks). 
While Ware~\cite{ware2012information} confirmed that redundancy encoding 
was integrated into the encoded dimension, in our case, where color and size were separable, we 
suggested that the redundancy worked because participants could use either length or color in 
different task conditions.
Since we could also consider that $length_y$-$color/length_x$ was a redundant encoding with $length_y$-$length_x$ and did
better than $length_y$-$color/length_x$ in some tasks (MAG and COMP),
we may arrive at a design recommendation: \textbf{when integral dimensions of $length_y$-$length_x$ were less accurate, adding more separable color 
could compensate to aid 
participants in their tasks. 
}
\remove{adding a more separable dimension to the 
integral encoding may help improve task completion time and accuracy.}

\subsection{Summary}


\add{
All tasks (MAG, RATIO, and COMP) lacked of significant main
effect on relative errors (in MAG or RATIO) or accuracy (in COMP).} 
Note that none of these three
tasks required initial visual search, and target answers were
labeled.
Wolfe called this type of task-driven with known target guided tasks{~\cite{wolfe2015guided}}. $Length$-$color$ was most accurate
in all task types. 
We thought at first that relative error may be
related to so-called proximity~\cite{wickens1995proximity}, i.e., the perceptual closeness
of visual variable choices to the tasks. The coloring was
perhaps more direct. However, since the participants read
those quantities as they commented,
we suspect that the reason
for not observing differences could well be their similarities
in mentally computing load. Since the search-space set-size was small, participants found that utilizing the object-level features was sufficient.  
\remove{When search-space set-size
increases for the MAX tasks, the search becomes time consuming and the length pairs would not be effective.}
We subsequently carried out the second experiment to increase the set size to the entire scene to study the scene guidance and correspondence errors.

\section{Empirical Study II: Global Scene-Features}



\add{
So far, we have validated the efficiency and effectiveness of the separable pairs only for simple tasks with a couple of items. 
The goal of the second experiment is to address the high-level hypothesis to quantify the benefits of separable feature-pairs for tasks in search 
spaces as large as the entire dataset of several hundreds items. 
}

\subsection{Overview}

\add{We  had  two  considerations in setting up  this  experiment.  The  first was a statement about  feature design relevant to the global holistic experience. If the vector field  contains one  object  at  a  time,  then  the  integral and separable dimensions and associated correspondence error may  explain   our  object  experience as we have shown in Experiment I. However, when the  binding problem is raised by  looking at multiple vectors,  it is possible that object binding does not occur  at the individual object-level but rather at the scene-level first, perhaps governed by global  gestalt  features.
}

\add{
The second consideration is relevant to the \textit{correspondence errors} when
managing holistic global viewing experiences that  may  not  be  significant when a  few  items  are  compared.  Generally, subjective reports from  the  first  study indicate that  $length$-$color$  and $length$-$texture$  show  similar  perceptual speed. For a feature to actually
\textit{guide} attention, Wolfe{~\cite{wolfe2018preattentive}} suggests that  a just-noticeable-difference for  that  feature is  not  sufficient and  one must  also look at feature distractors, whether or not  they  are heterogeneous, and that the efficiency  of a scene guidance will decline  as a function of the degree of distractor variation. Efficiency  will  be  achieved if the  target and distractors are ``linearly separable'' meaning that a line can be drawn separating the  target  from  the  distractors in the feature space. This is similar  to the studies of Acevedo et al.{~\cite{acevedo2007modeling}} for saliency  measures and Urness et al.{~\cite{urness2003effectively}} for ``texture stitching''. Acevedo et al. attempted to show that features can be segmented and Urness et al show boundaries from continuous flow fields using spatial frequency. Chung et al. showed the order and just noticeble differences of visual stimuli and suggested that size and hue had highest accuracy in 2D and texture value introduces greater ordering effects~\cite{chung2016ordered}.
In our study, we believe that colors (especially in multiple hues) are categorical and thus should produce better  perceptual speed than texture and  length. Performance of  texture may decline faster than  color  as  the  data  range  increases because our vision  is not as sensitive to luminance-variation as to hues. The  efficiency   of color in Experiment I could   well  arise  because the  range  (of 4) was  not  large  enough. The  current study expands the  data  range   from  the  single  level  in  the  first study to  five  ranges $\in[3, 7]$ to  understand feature-pair  scalability. SplitVectors produces the \textit{second type of correspondence error} between two lengths which can challenge human eyes to see the two component parts.
}

\subsection{Method}

\subsubsection{Feature-Pairs}

\begin{figure}[!tb]
\centering

\begin{subfigure}[b]{\columnwidth}
		\centering
		\includegraphics[width=0.75\columnwidth]{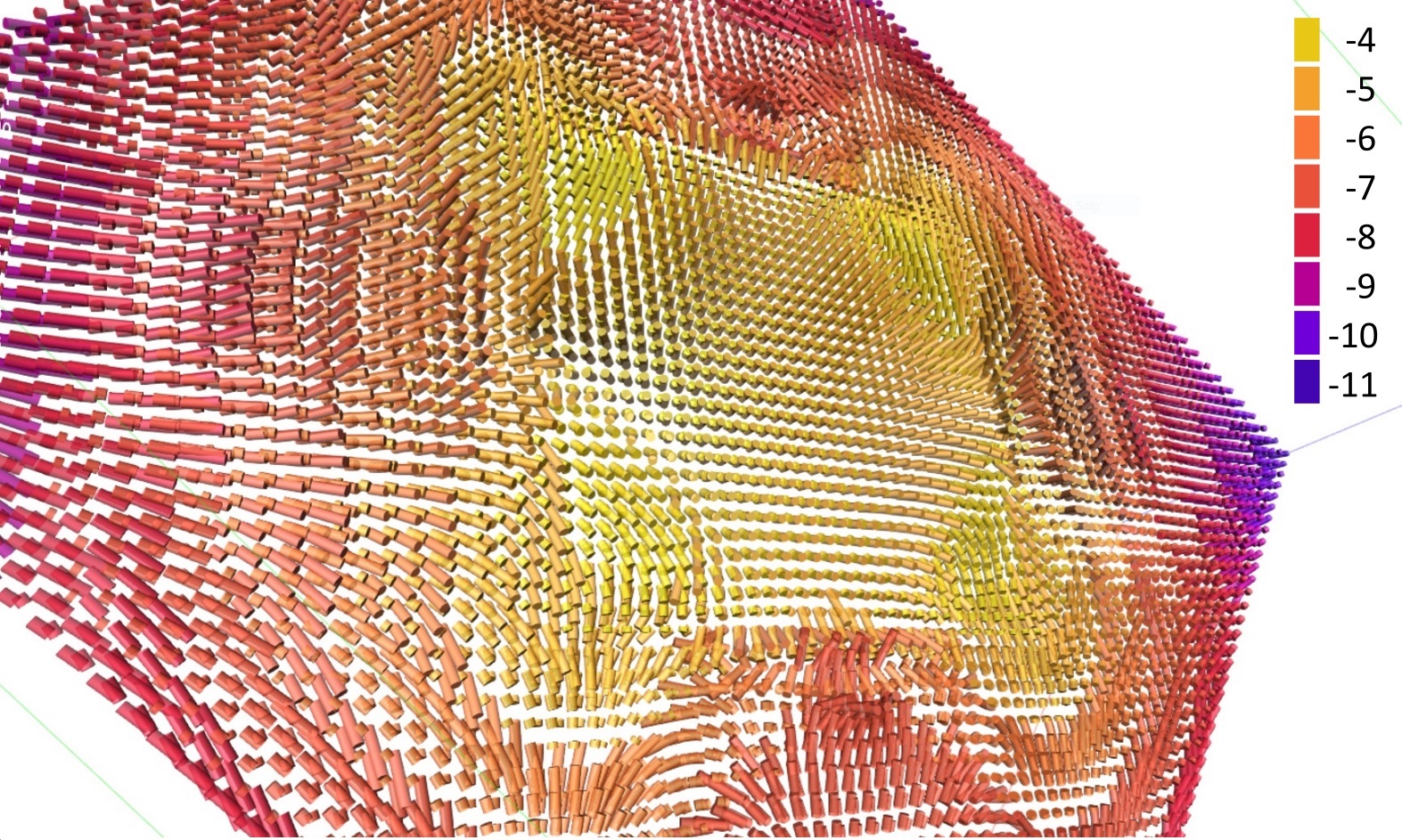}
\caption{Continuous colormap and high-density data}
\label{fig:continousHigh}
\end{subfigure}

\begin{subfigure}[b]{\columnwidth}
	\centering
	\includegraphics[width=0.75\columnwidth]{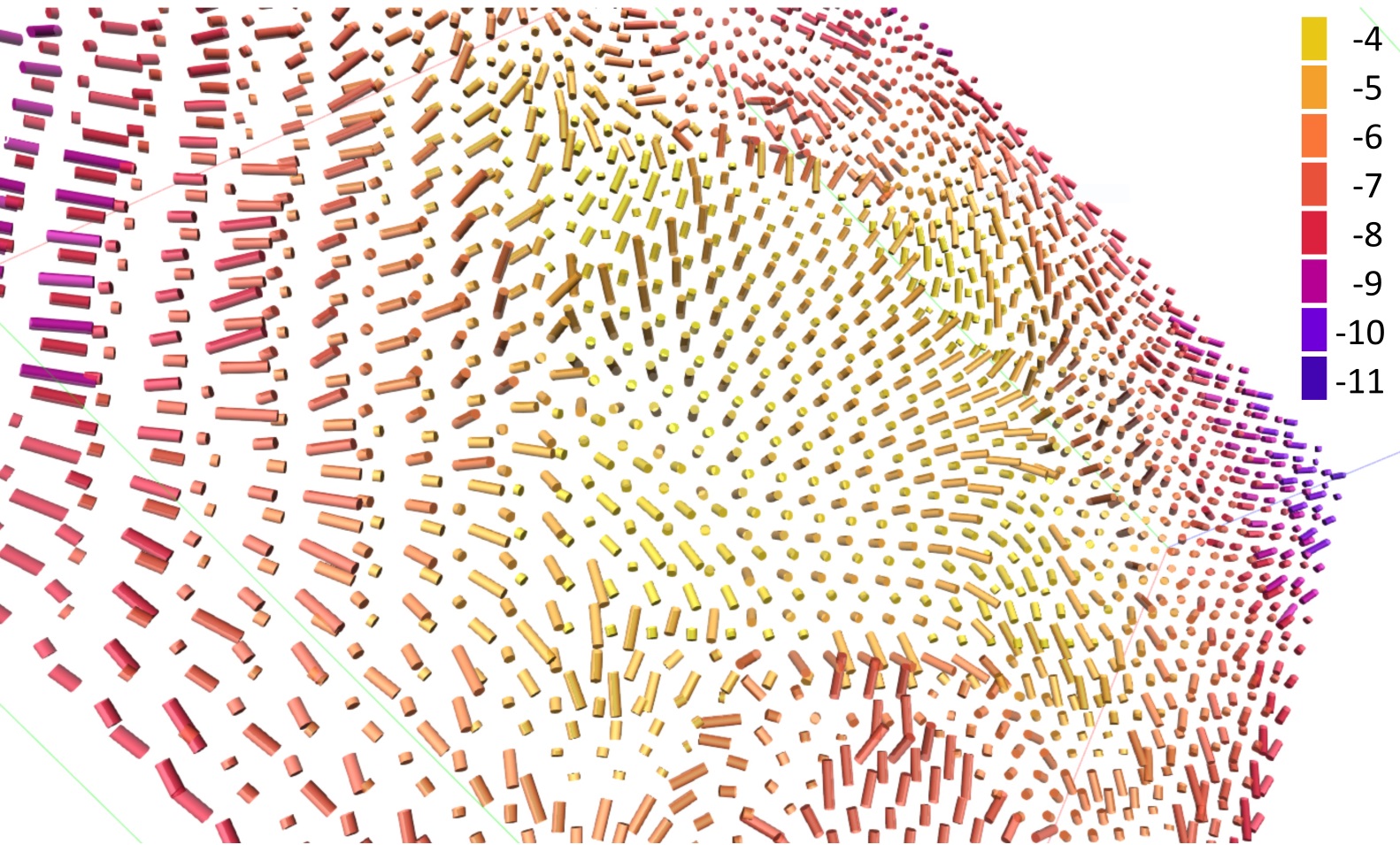}
		\caption{Continuous colormap and low-density data}
\label{fig:continousLow}
\end{subfigure} 

\begin{subfigure}[b]{\columnwidth}
		\centering
\includegraphics[width=0.75\columnwidth]{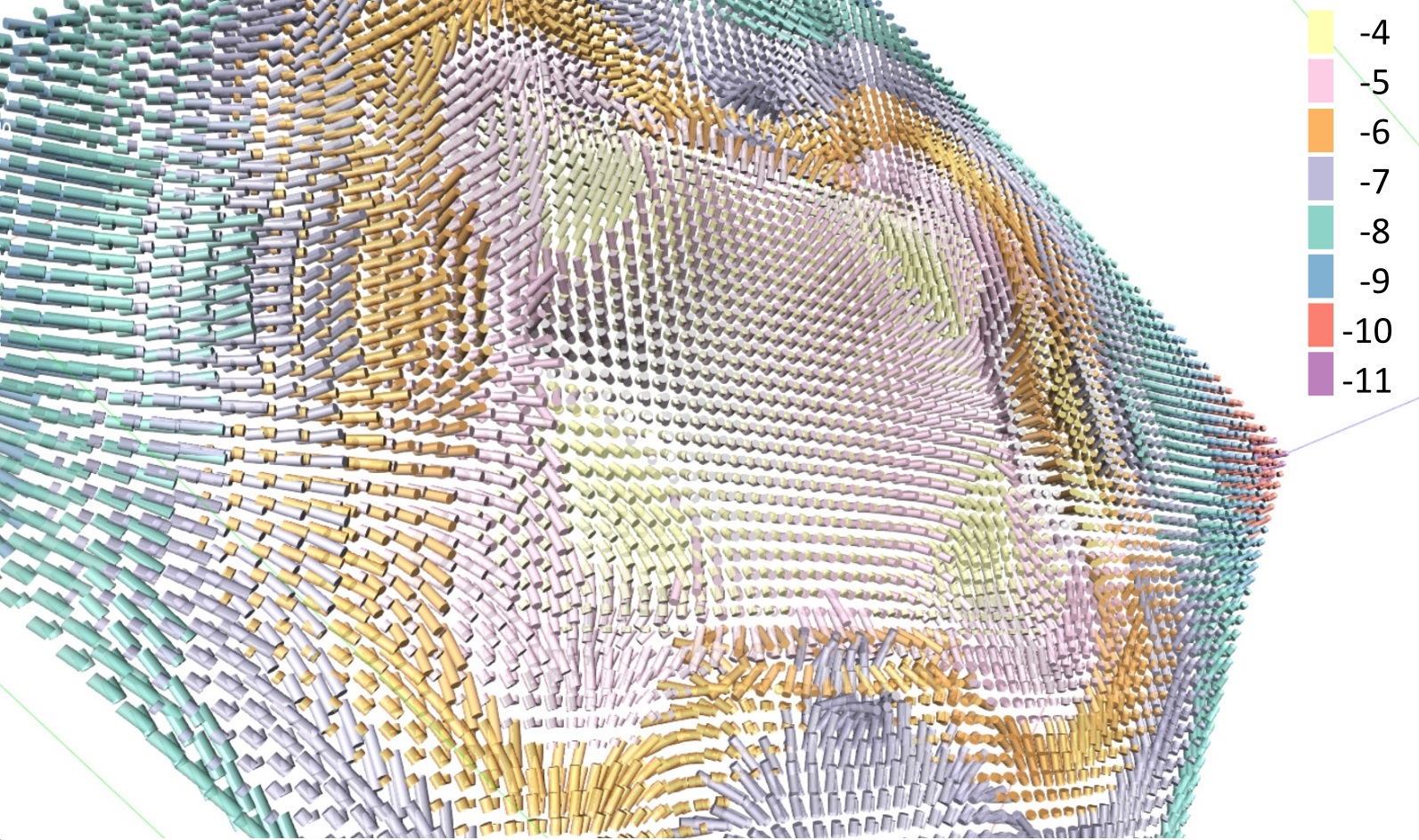}
		\caption{Categorical colormap and high-density data}
		\label{fig:catogoricalHigh}
\end{subfigure} 

\begin{subfigure}[b]{\columnwidth}
	\centering
\includegraphics[width=0.75\columnwidth]{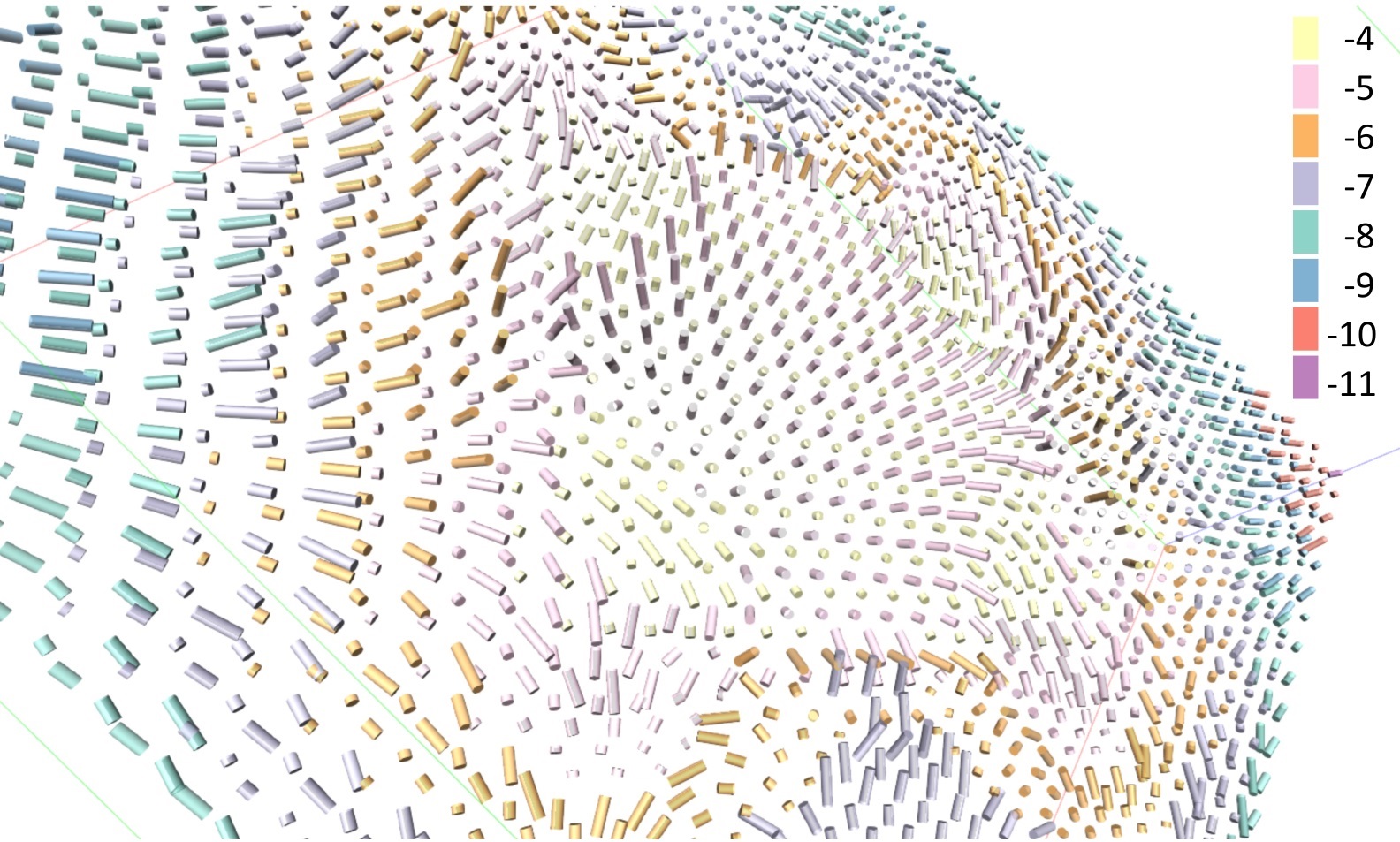}
\caption{Categorical colormap and low-density data}
\label{fig:catogoricalLow}
\end{subfigure}
\caption{Experiment II: An example data using a categorical and a segmented continuous colormaps with two data densities. The boundaries between the data categories are more recognizable when the data are dense in (a) and (c). The boundaries are more difficult to recognize in (b) than in (d). We use the categorical colormaps in Experiment II. 
}
\label{fig:colorAndData}
\end{figure}

\add{
We used $length$-$color$, $length$-$texture$, and baseline splitVectors in Experiment II. 
These three visualizations were chosen because $length$-$color$ and $length$-$texture$ are among the best feature-pairs from Experiment I and because color and texture are among the most separable features according to Ware{~\cite{ware2012information}}.
To introduce a \textit{correspondence error} or ``distractor'' experience, we vary the data range from the 4 levels in experiment I to 3-7 levels in Experiment II. 
}

\add{
We had two reasons to use categorical hue
instead of quantitative colormaps. The first was based on the subjective observation comparing a categorical colormap from Colorbrewer~\cite{harrower2003colorbrewer} and a segmented continous colormap by the number of exponents generated from the extended blackbody colormap (Figure{~\ref{fig:colorAndData}}). As we can see, the boundary detection with these colormaps might be associated with data density.
We found that unless the data density was reasonably high, detecting the boundaries using continous colormaps (Figures{~\ref{fig:continousHigh}}, {~\ref{fig:continousLow}}) was harder than the Colorbrewer colormaps (Figures{~\ref{fig:catogoricalHigh}}, {~\ref{fig:catogoricalLow}}). The second reason is that the initial \textit{at-a-glance} global statistical summary of the scene depends on categorical information~\cite{wolfe2018preattentive} - then categorical visual encoding may be more suitable. An informal observation on texture choices was that detecting maximum and minimum regions was easier than any intermediate regions.
}

\subsubsection{Hypotheses}

We had the following hypotheses:

\begin{itemize}
\item
\textit{Exp II.H1. (Accuracy). More separable pairs will be more effective. We thus  anticipate a rank order of effectiveness from high to low: $length$-$color$, $length$-$texture$, and $splitVectors$.}

While we did not see a significant main effect in Experiment I and we believed that for tasks related to multiple objects (vectors),
pop-out color features would reduce the correspondance error and facilitate scene-level feature binding.

\item
\textit{Exp II.H2. (Correspondence error)}.
\textit{Less separable pairs would lead to 
more first type correspondence errors, when participants would choose the wrong exponent level.}

\item 
\textit{Exp II.H3. (User behavior)}. 
\textit{More separable feature-pairs would lead to optimal
users' behaviors: i.e., participants can quickly locate 
task-related regions for tasks that demand looking
among many vectors.} 
\end{itemize}


\subsubsection{Tasks}

\begin{figure}[tb]

          \begin{subfigure}[t]{0.4\textwidth}
        		\centering
        		\begin{tikzpicture}
        		\node (img1){\includegraphics[width=\textwidth]{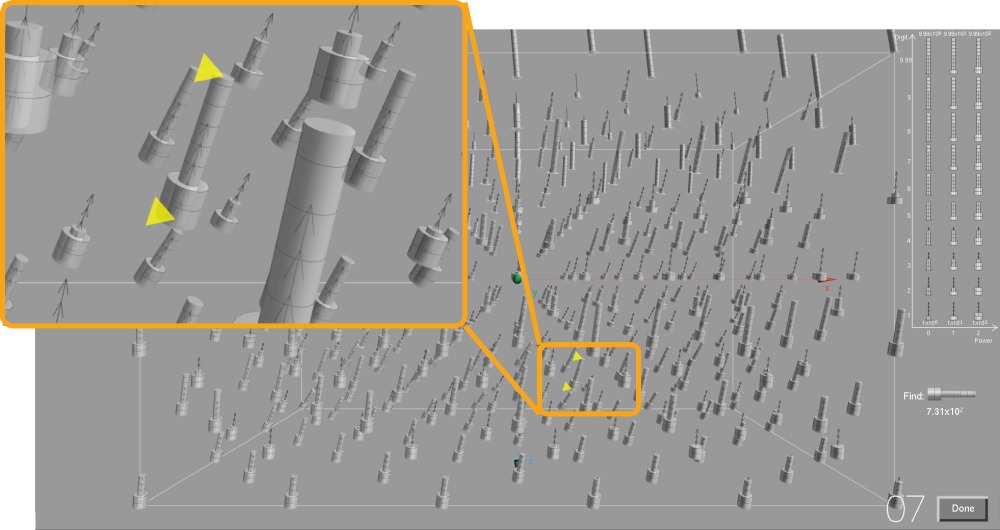}}; 
        		\end{tikzpicture}
        		\caption{SEARCH: Find the vector with magnitude X. (X: 731, answer: the point marked by two yellow triangles. No answer or feedback was provided during the study.)}
        		\label{fig:SEARCH}
        	\end{subfigure}

			\begin{subfigure}[t]{0.4\textwidth}
        		\centering
        		\begin{tikzpicture}
        		\node (img1){\includegraphics[width=\textwidth]{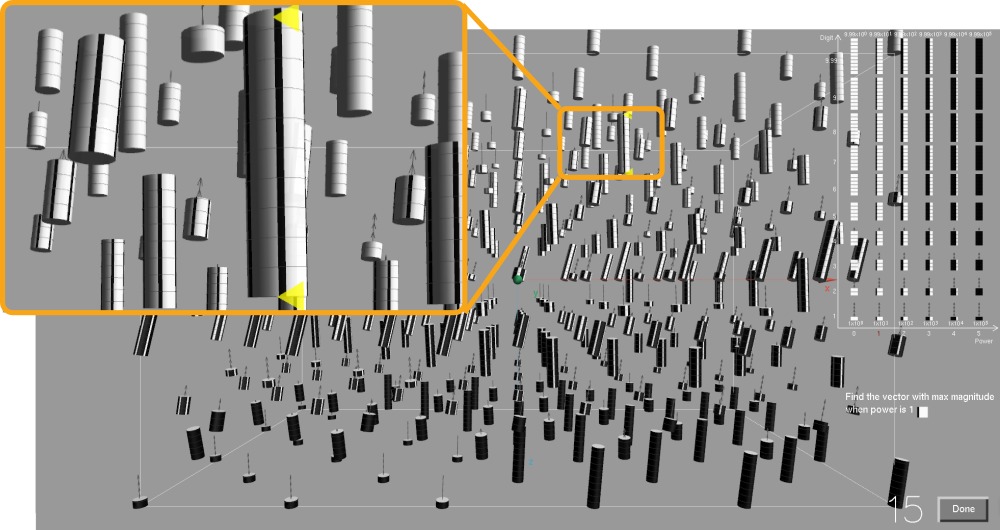}}; 
        		\end{tikzpicture}
        		\caption{MAX: Which point has the maximum magnitude when the exponent is X? (X=1, answer: the point marked by two yellow triangles. No answer or feedback was provided during the study.)}
        		\label{fig:MAX}
        	\end{subfigure}       

          \begin{subfigure}[t]{0.4\textwidth}
        		\centering
        		\begin{tikzpicture}
        		\node (img1){\includegraphics[width=\textwidth]{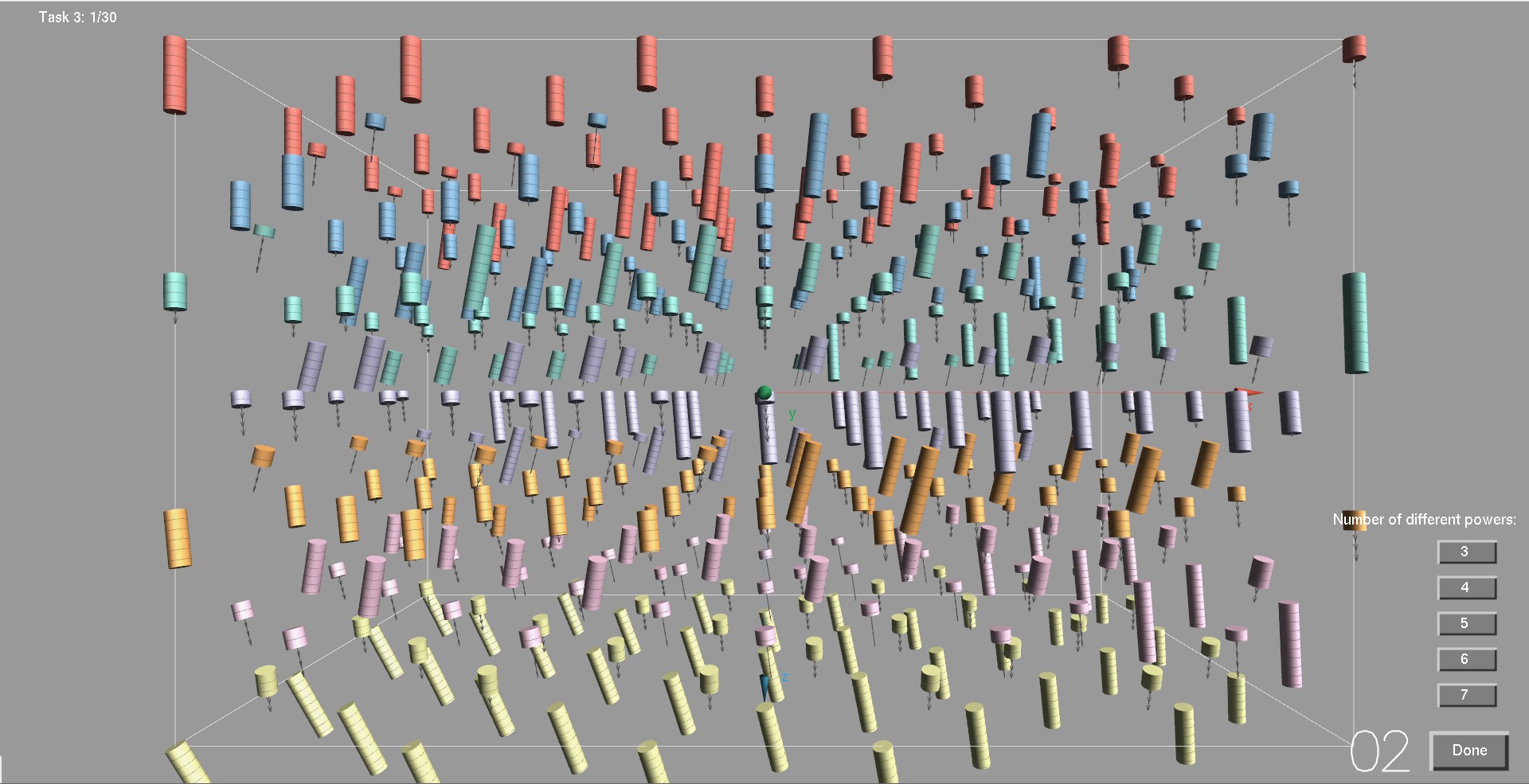}}; 
        		\end{tikzpicture}
        		\caption{NUMEROSITY (NUM): Estimate the total number of vector exponents of the entire vector field within 2 seconds. (answer: 7)}
        		\label{fig:NUM}
        	\end{subfigure}
        	
			 \caption{Experiment II three task types. The callouts show the task-relevant feature-pair(s).}
             
             \label{fig:exp2tasks}
        \end{figure}

Participants performed three tasks in which they had to compare all vectors to obtain an answer.

\textbf{Exp II.Task 1 (SEARCH): A vector search within 20 seconds  (Figure~\ref{fig:SEARCH}).} \textit{Find the vector with magnitude X within 20 seconds.}
The target vector was shown at the bottom-right corner of the screen. Participants were asked to find this vector.

\textbf{Exp II.Task 2 (MAX):  An  extreme value search within 20 seconds (Figure{~\ref{fig:MAX}})}.
\textit{Within 20 seconds, locate the point has maximum magnitude when the exponent is X.}
X in the study was a number from 0 to the \textit{maximum exponent} ($\in[2, 6]$). 
This was a global task requiring participants to find the extremum among many vectors.

\textbf{Exp II.Task 3 (NUMEROSITY): within 2 seconds, estimate the total number of vector exponents (Figure~\ref{fig:NUM})}. \textit{Estimate the total number of vector exponents in the entire vector field within 2 seconds.}
Data are randomly chosen and modified to produce the 3 to 7 range. No data is used repeatedly in this experiment.

\subsubsection{Data Choices}

\add{
Data were  first sampled using  the same  approach as Experiment I. We then  modified the exponent range from 3 to 7 for the three  tasks by normalizing the data  to the desired new  data  range.  Doing  this let us  preserve the  critical  domain-specific data  attributes of their  spatial structures and  only altered the magnitude range  to improve the applicability and  reuse  of our study results.
}

Prior  literature used  both  synthetic data  and  real-world data  to  construct the  data  visualization as  test  scenarios, enabling tight  control  over  the  stimulus parameters. Most of  the  synthetic data   in  these  studies were   generated to replicate real-world data  characteristics and  others were explained in  fictitious use scenarios.  The goal was primarily to prevent preconceived user knowledge about the  domain-specific attributes. As a result,  the  synthetic data  strike  the  right  balance  between real-world uses  and the  data  characteristics. In our  cases,  replicating characteristics in quantum physics  data  was  challenging and indeed  impossible, since atom  behaviors in  high-dimensional space  were  largely  unknown and thus  were  not  easily  simulated. Our  approach was  therefore to  randomly sample quantum physics  simulation results to capture domain-specific  attributes and  then  modify the  data  to  suit  evaluation purposes. We  showed our  data  to  our physicist collaborators to ensure their  validity.




\subsubsection{Empirical Study Design}
\emph{Dependent and Independent Variables.}
We used a within-subject design with two  independent  variables of 
\textit{feature-pair} (three levels: baseline splitVectors, $length$-$color$, and $length$-$texture$) and \textit{exponent range} (five levels: 3-7). 
The dependent variable was  relative error.  We did  not measure time since all tasks  were  time-constrained.

Participants performed 3 (feature-pairs) $\times$ 5 (magnitude-ranges) = 15 trials for the first two tasks. Three repetitions were used to give participants enough time to develop strategies. 
For NUMEROSITY tasks, the design runs 4 repetitions, resulting in 3 (feature-pair) $\times$ 5 (exponent-range) $\times$ 4 (repetition) = 60 trials. Each participant thus executed $45+45+60 = 150$ trials. Completing all tasks took about 32 minutes.

\emph{Self-Reporting Strategies.} 
Several  human-computer interaction   (HCI)  approaches can  help   observe users'  behaviors.  Answering questions can  assist  us  to  determine not just  which  technique is better  but  also  the strategies humans adopt. For  example, cognitive walkthrough  (CTW)  measures  whether or  not  the  users’  actions  match  the  designers'  pre-designed  steps.   Here   we  predicted  that   participants would use  the  global  scene-features as  guidance to accomplish tasks. We interviewed participants and  asked them  to verbalize their  visual  observations in accomplishing tasks.

\subsubsection{Participants}
Eighteen new  participants (12 male and 6 female, 
mean age = 23.8, and standard deviation = 4.94) of diverse backgrounds participated in the  study (seven in computer science, four in computer engineering, two in information systems, three in engineering, one in business school, and one in physics).
Procedure, interaction, and  environment were  the same  as those in the Experiment I.

\subsection{Experiment II: Results and Discussion}

We collected 810 data points per task for the first two tasks of SEARCH and MAX and 1080 points for the third NUMEROSITY task. 

\subsubsection{Summary Statistics}


\begin{table}[!tb]
	\caption{Experiment II: Summary statistics by tasks. The significant main effects and the high effect size are in \textbf{bold} and the medium effect size is in \textit{italic}. Effect size is 
	Cohen's d for tasks SEARCH and MAX, and Cramer's V for task NUMEROSITY (NUM). Post-hoc Tukey grouping results are reported for significant main effects, where $>$ means statistically significantly better and enclosing parentheses mean they belong to the same Tukey group. Here, LC: $length$-$color$ and LT: $length$-$texture$.}
	\label{tab:exp2result18}
	\begin{center}
		\begin{tabular}{l l l l}
		\toprule
\centering Task & Variables & Significance & ES\\
			\midrule
    	SEARCH & feature-pair & \textbf{F$_{(2,\,261)}$ = 18.4}, \textbf{\textit{p} $<$ 0.0001} & \textbf{0.46}\\ 
    	& & \textbf{(LC, LT)} $>$ \textbf{splitVectors}& \\
		& power-range & F$_{(4,\,261)}$ = 1.5, p = 0.20  & \textbf{0.86}\\
			\midrule	
	MAX & feature-pair & \textbf{F$_{(2,\,261)}$ = 15.8}, \textbf{\textit{p} $<$ 0.0001} & \textbf{0.47} \\ 
	& & \textbf{(LC, LT)} $>$ \textbf{splitVectors}& \\
		& power-range & F$_{(4,\,261)}$ = 0.3, \textit{p} = 0.87 & 0.11\\
			\midrule
	NUM & feature-pair &  \textbf{${\chi}^2$ = 63.2}, \textbf{\textit{p} $<$ 0.0001} & \textit{0.25} \\
	& & \textbf{LC} $>$ \textbf{splitVectors} $>$ \textbf{LT}& \\
   & power-range &  \textbf{${\chi}^2$ = 47.4}, \textbf{\textit{p} $<$ 0.0001}  &  \textbf{0.35} \\
   & &  \textbf{(3, 4)} $>$ \textbf{5} $>$ \textbf{(6, 7)}&\\
			\bottomrule
		\end{tabular}
	\end{center}
\end{table}

\begin{figure}[!tb]
\centering
	\begin{subfigure}[t]{0.48\textwidth} 
        	\includegraphics[width=\textwidth]{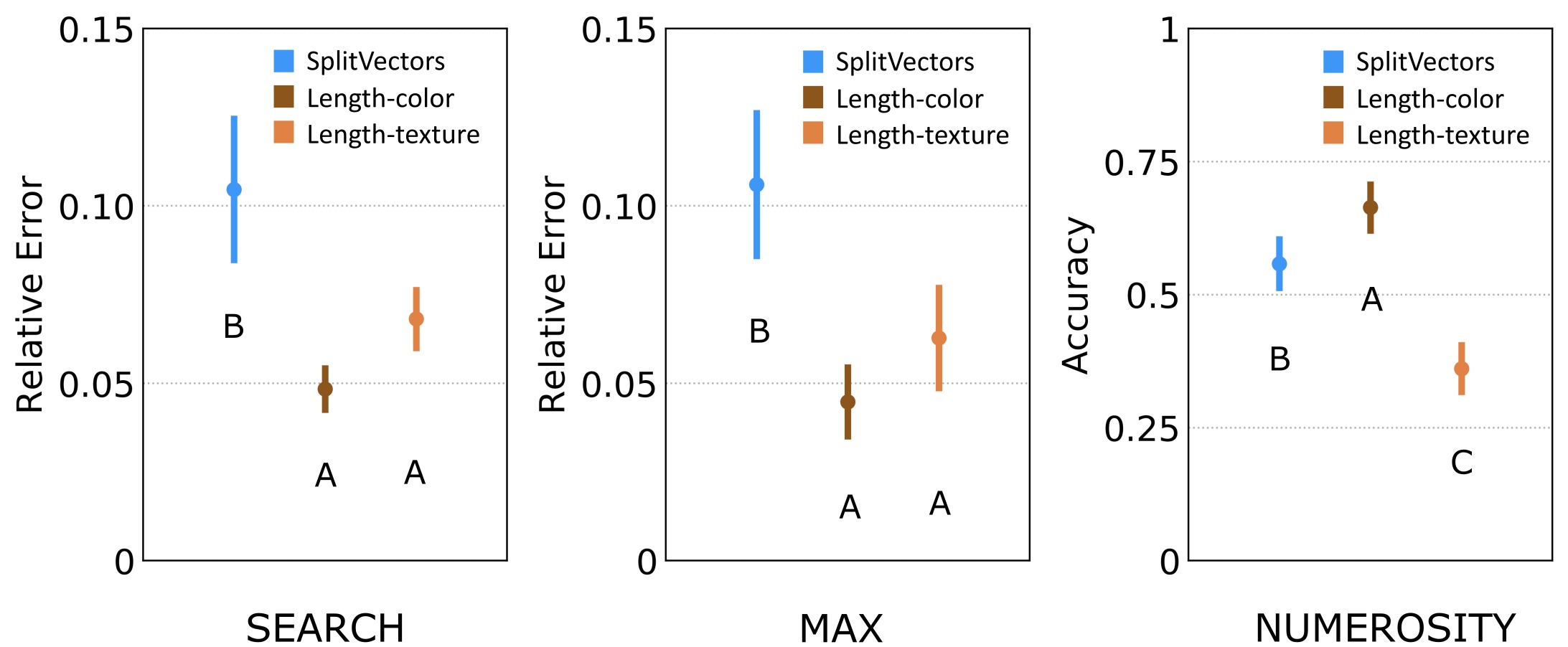}
		\label{fig:exp2max_feature}
	\end{subfigure}
	\begin{subfigure}[t]{0.48\textwidth}
        	\includegraphics[width=\textwidth]{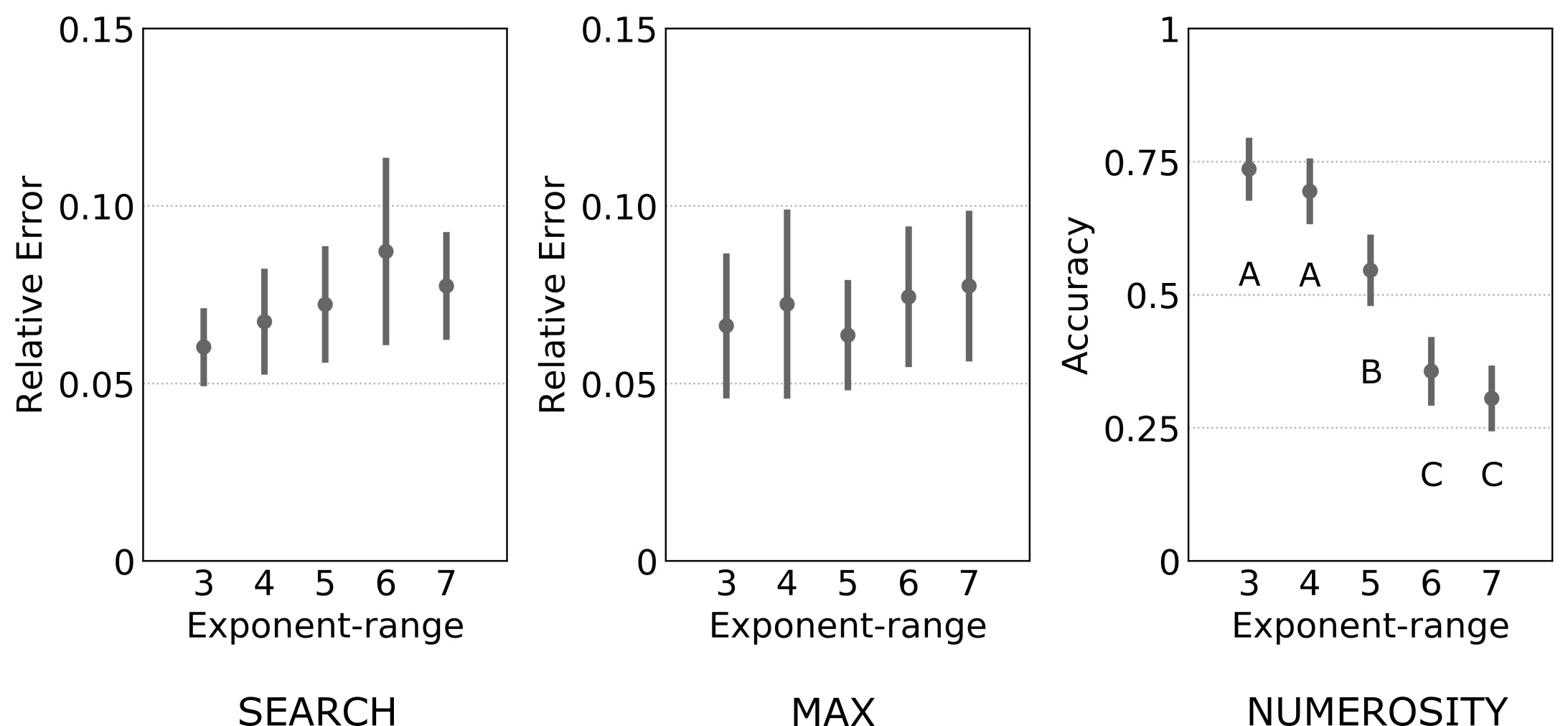}
		\label{fig:exp2num_feature}
	\end{subfigure}

\caption{Experiment II: Relative error in SEARCH and MAX and accuracy in NUM.
Same letters represent the same post-hoc analysis group. All error bars represent $95\%$ confidence intervals.
}
\label{fig:exp2result}
\end{figure}

For SEARCH and MAX tasks, we measured relative error (which was the percentage the reported value was away from the ground truth) with SAS repeated measure for SEARCH and MAX.
The last NUMEROSITY tasks used accuracy which was the percentage of correct answers of all trials for each participant.
Table~\ref{tab:exp2result18} and Figure~\ref{fig:exp2result} show the summary statistics; And all error bars again represent $95\%$ confidence intervals.

We observed a significant main effect of feature-pair type on all three tasks. 
For the first two tasks, post-hoc analysis revealed that $length$-$color$ and $length$-$texture$ were in the same group, the most efficient one and that relative errors were statistically significantly lower than those of the splitVectors. $Length$-$color$ remains the most accurate pair for the NUMEROSITY tasks. 
Exponent-range was only a significant main effect for NUMEROSITY, with power ranges 3 and 4 were significantly better than 5, which was better than 6 and 7.

\subsubsection{More Separable Dimensions Improved Accuracy}

\add{
We  were  interested to see if we  could   observe significant main effects for features which showed none in local comparison tasks  in  Experiment I. Here  we did observe the significant main effect and confirmed our  first hypothesis: $length$-$color$  and  $length$-$texture$ were  also  in  the  same 
more efficient Tukey group for both  SEARCH  and MAX. 
All participants reported  that they   searched   the   exponent  terms  first  and then the digits following our hypothesis that there may exist a global pop-out of the \textit{color} and \textit{texture} features. Participants preferred colors and they could easily differentiate the powers.
}

\add{
$Length$-$color$  also  led  to  the most  accurate answers, and 
now \textit{splitVectors} was  better  than   $length$-$texture$ for NUMEROSITY
tasks. This result can be explained by participants' behaviors
- more  than  half the  participants suggested they simply look for the longest cylinder from the splitVectors since they know  the numerical values in the test were continuous. This behavior deviated from  our  original purpose of testing the global estimate but did show  two perspectives in favor of this  work:  (1) participants developed task-specific strategies during the experiment for efficiency;  (2) 3D length still supported reasonable pop-out feature, even though it was not as effective as color.
}


\add{
These subjective behaviors through self-report suggested that they
adopted a
sequential task-driven viewing strategy to first obtain gross regional distribution
of task-relevant exponents. After this, a visual comparison within the same exponent region were achieved; And no
binding at the object-level was performed especially when the features pop-out globally as scene features. With these
two steps, judging large or small or perceiving quantities
accurately from separable variables  would not use object-binding.
Participants in our study used a top-down control process to utilize these spatial constraints regardless of feature types (for NUMEROSITY) and modify how global structures are used to see the features between tasks.
}

\subsubsection{The Cost of Correspondence Errors}

\add{
Reducing correspondence error was influenced by the choices of separable dimensions.
} Our  second hypothesis H2  (correspondence error)  was  also  supported. We  first tested  the first type of correspondence error (those answers with  different exponent values) in  MAX  and SEARCH  in the same way as in Experiment I. We saw 36 instances from SEARCH (about $4.4\%$ of all samples, or  1 $length$-$color$, 20 $length$-$texture$, and  15 \textit{splitVectors}); 59 instances from  MAX (about $7\%$ of all samples or  6 $length$-$color$, 38 $length$-$texture$, and  15 $splitVectors$). 
These results when combined with those in Experiment I confirmed that  $length$-$texture$ had worse first type correspondence error. However when viewers in the correct data sub-categories, they could obtain as accurate answers as $length$-$color$.

\add{
All participants commented on how the number of powers in the data affected their effectiveness. For  $length$-$texture$, 10 participants remarked that  it was  difficult to differentiate adjacent powers when the total power level is around 4-5 for $length$-$texture$. The  white  and  black textures  were   very  easy  to  perceive. All but two participants agreed that  $length$-$color$  could  perhaps support up  to  6.  Chung et al.{~\cite{chung2016ordered}} studied ordering effects and it would be challenging to compare ours to their results because their visual stimuli were not shown as a scene-feature but a stimuli alone. 
More  than  half  of  the participants felt  that  effectiveness of splitVectors  was not affected  by  changing the  number of powers, since they  looked  for  the  longest outer cylinder to  help  find  the answer.
These results may suggest that subregion selection with $length$-$texture$ can perhaps be better designed with interfaces when the users can interactively select a texture level. 
}



\section{General Discussion}

We discuss the results from both experiments and suggest future directions.

\subsection{Separable Dimensions for Univariate Data Visualization for Large-Range Quantum Physics Data}

\add{
The  results  of  Experiment I  showed  that  separable dimensions could achieve the same efficiency as direct linear visualizations for the COMP task and was always more efficient than integral pairs. For these local-tasks, we didn't observe significant error reduction.
The results from Experiment II studied the rank order of the separable pairs and found that  more separable pairs also improved accuracy for global tasks. $length$-$texture$ and splitVectors in both experiments led to higher correspondence errors than $length$-$color$.
}

Visual   variables  that   are   separable  (i.e.  manipulated and  perceived independently) would initially be considered problematic  for   encoding  univariate  data   because  of the known object-level feature-binding challenges involving in achieving integrated numerical readings by combining two visual features.  
Our experiment showed that binding does not have to be successful at the object-level. A viewer  can adopt a sequential task-driven viewing strategy based on a view hierarchy: viewers first  obtain  global  distributions of the scene. Then,  a visual scrutiny is possible within a subregion. In other words, binding occurred at the  scene level rather than  the object level.


\add{
The separable-dimension pairs of $length$-$color$ and $length$-$texture$ worked  because they supported the scene-centered structural perception in which  the processing of global structure and the spatial relationships among components precede analysis of local details according to participants' self-reports.
Another possibility for texture to be effective is the ordering - participants could see large and small{~\cite{chung2016ordered}}. From  a practical perspective, our  results may suggest that it was easiest for viewers to  interpret a  scene in which features are scene features for showing ordering and global structures. Scientific  data  are  rarely unstructured. Using  coloring to provide some initial regional division may  be always better than not. Texture (luminance) could achieve similar accuracy and efficiency as long as the first-type of correspondence error was removed. 
}

\subsection{Feature Guidance vs. Scene Guidance}

\input{figureCase2.tex}

Taking into account both study results, we think an important part of the answer to \textit{correspondence error} is \textit{guidance} of attention.  Attention in most task-driven fashion is not deployed randomly to objects.  It is guided to some objects/locations over others by two broad methods:  \textit{feature guidance} and \textit{scene guidance}. 

Feature guidance refers to guidance by properties of the task-target as well as the distractors (leading to correspondence errors). These features are limited to a relatively small subset of visual dimensions: color, size, texture,  
orientation, shape, blur or shininess and so on.
These features have been broadly studied in 3D glyph  design (see reviews by Healey  and  Enns~\cite{healey2012attention}, Borgo  et al.~\cite{borgo2013glyph}, Lie et al.~\cite{lie2009critical},  Ropinski et al.~\cite{ropinski2011survey}, and McNabb and Laramee~\cite{mcnabb2017survey}).
Take one more example from quantum physics simulation results, but with a different task of searching for the structural distributions  in the power of 3 in Figure~\ref{fig:case2} will guide attention to either the fat cylinders (Figure~\ref{fig:case2LLO}) or the bright yellow color (Figure~\ref{fig:case2LC}, ~\ref{fig:case2LCL}) or the very dark texture (Figure~\ref{fig:case2LT}), depending on the feature-pair types. 

Working with quantum physicists, we have noticed that the structure and content of the scene strongly constrain the possible location of meaningful structures, guided ``scene guidance'' constraints~\cite{biederman1977processing, wolfe2015guided}.
Scientific data are
not random and are typically structured. 
Contextual and   global   structural  influences can  arise from  different sources of  visual   information. 
If we return to 
the MAX search task in Figure~\ref{fig:case2} again, we will note that
the chunk of darker or lighter texture patterns and colors
on these regular contour structures strongly influence our
quick detection. This is a structural and physical constraint
that can be utilized effectively by viewers. This observation
coupled with the empirical study results may suggest an
interesting future work and hypothesis: 
\textbf{adding scene structure guidance
would speed up quantitative discrimination, improve the
accuracy of comparison tasks, and reduce the perceived
data complexity.}

Another structure acting as guidance is the size itself. It was used  by  participants seeking  to  resolve  the  NUMEROSTIY tasks to look for the  longest outside cylinders.
We have showed several examples like Figure~\ref{fig:case2}, our  collaborator suggested that the cylinder-bases of the same size with the redundant encoding (Figure~\ref{fig:case2LCL}) also helped locate  and group glyphs belonging to the same magnitude.
This observation agrees with the most recent literature that
guidance-by-size in 3D must take advantage of knowledge
of the layout of the scene~\cite{eckstein2017humans}.

Though feature guidance can  be  preattentive and  features   are  detected  within  a  fraction  of  a  second,  scene guidance is probably just about  as fast (though precise experiments have not been done and our Experiment II only merely shows this effect). Scene `gist' can be extracted from  complex images  after  very  brief  exposures~\cite{biederman1977processing, oliva2005gist}. This doesn't  mean  that  a viewer  instantly knows, say,  where the answer is located.  However, with a fraction of a second's  exposure, a viewer  will know enough about  the spatial layout of the scene to guide his or her attention towards vector groups in the regions of interest.

\add{
A future direction, and  also an  approach to understanding the  efficiency  and  the  effectiveness of scene  guidance, is to conduct an eye-tracking study to give viewers a flash-view of our spatial structures and  then  let  the  viewer  see  the display only in a narrow range  around the point  of fixation: does  this  brief  preview guide attention and  the gaze effectively?  Recently,  work in the information visualization{~\cite{ryan2019glance}~\cite{bylinskii2015intrinsic}~\cite{borkin2013makes}} domain has  measured and  correlated performance on  the glance  or  global  structure  formation. Vision  science  discovered long ago that  seeing  global  scene structures in
medical imaging decision making guides experts’ attention (experts always know  where to look){~\cite{kundel2007holistic}~\cite{drew2013invisible}}.
}

\remove{
In grammar of Graphics Wilkinson puts forward some plausible properties that `nice' scales should possess and suggests a possible algorithm. The properties (simplicity, granularity and coverage, with the bonus of being called `really nice' if zero is included) are good but the algorithm is easy to outwit.  Difficult cases for scaling algorithms arise when data cross natural boundaries, e.g., data with a range of 4 to 95 would be easier to scale compared to 4 to 101. 
}



\subsubsection{Use Our Results in Visualization Tools and Limitations of Our Work}

Visualization is used when the goal is to augment human capabilities in situations where the problems might not be sufficiently defined for a computer to handle algorithmically or to communicate certain information.
One of these areas is quantum physics: simulation results are in high-dimensional space thus cannot be interpreted in computational solutions. As a result, quantum physicists count on visualization to detect patterns and trends. Our collaborators were amazed though not surprised by many design possibilities and the performance differences among them.  

Our current study concerns bivariate data visualization in which the bivariate variables are component parts of a univariate variable. the first variable is 
always an integer and the second variable is bounded to a
real number in the range [1, 10). Application domains carrying similar data attributes could reuse of work. 
The design principle of prompting scene-level guidance would be broadly applicable to 3D visualizations. 
Our design is somewhat limited to preliminary pop-out stimuli. Our design could have been improved by following advanced tensor glyph design methods especially those in tensor field visualizations. Both generic{~\cite{gerrits2017glyphs}}  and domain-specific requirements for glyph designs {~\cite{zhang2016glyph}~\cite{schulz2013design}~\cite{kindlmann2006diffusion}} have led to the summary of glyph propertise (e.g., invariance, uniqueness, continuity) to guide design and to render 2D and 3D tensors. A logic step for us is to truly understand the quantum physics principles to combine data attributes and human perception to arrive domain-specific solutions.  

One limitation of this work is that we measured only a subset of tasks crucial to showing structures and omitted all
tasks relevant to orientation. However, one may argue that the vectors naturally encode orientation. 
When orientation is considered, we could address the multiple-channel mappings in two ways.
The first solution is to use the $length$-$texture$ to encode the
quantitative glyphs and color to encode the orientations if we cluster the vectors by orientations.
The second solution is to treat magnitude and orientation as two
data facets and use multiple views to display them separately, with one view showing magnitude and the
other for orientation (using Munzner's multiform design recommendations~\cite{munzner2014book}).
The second limitation here was that our experiments were limited to  a relatively small  subset  of visual  dimensions: color, texture, and size. A future direction would be to try shapes and glyphs to produce novel and useful design.

\section{Conclusion}


This work  shows  that  correspondence computation is necessary  for retrieving information visually and  that  viewers’ strategies can  play  an  important role.  Our  results showed that  $length$-$color$ with the separable pairs
\change{fall into the same group as the  linear  ones.}
{was most efficient and effective for both local and global tasks.}
Our  findings in general suggest that, as we hypothesized, distinguishable separable dimensions perform better.  Our   empirical study  results provide the following recommendations for  designing  3D bivariate glyphs \add{for representing univariate variables}.

\begin{itemize}
\item
Highly separable pairs can be used for quantitative \remove{holistic data} comparisons as long as these glyphs are \add{scene-}structure forming. 
We recommend using $length$-$color$. \remove{ and $length_y-texture$.}
\item
Texture-based glyphs ($length-texture$) that introduces \change{spatial-frequency}{luminance} variation \change{are recommended.}{can cause correspondence error and will only be recommended when task-relevant structures can be constained.}
\item
Integral and separable bivariate \change{glyphs}{feature-pairs} have similar accuracy when the tasks are
\change{guided (aka, target location is known).}{local.} They influence  
accuracy \remove{only when the target is  unknown and} when the search space increases. 
\item
3D glyph scene would shorten task completion time when the glyph scene support structural feature guidances.
\item
The redundant encoding ($length_y$-$color/length_x$) greatly improved on \change{the performance}{task completion time} of integral dimensions (splitVectors) by adding separable and preattentive color features.
\end{itemize}

Empirical study data and results can be found online at
\textit{https://sites.google.com/site/interactivevisualcomputinglab/download /integral-and-separable-dimension-pairs}.
\ifCLASSOPTIONcompsoc
  \section*{Acknowledgments}
  The work is supported in part by NSF IIS-1302755, NSF CNS-1531491, and NIST-70NANB13H181.
The user study was funded by NSF grants with
the OSU IRB approval number 2018B0080. 
Non-User Study design work was supported by grant from NIST-70NANB13H181. 
The authors would like to thank Katrina Avery for her excellent editorial support 
and all participants for their time and contributions. 

Any opinions, findings, and conclusions or recommendations expressed in this material are those of the author(s)
and do not necessarily reflect the views of the National Science Foundation. Certain commercial products are
identified in this paper in order to specify the experimental procedure adequately. Such identification is not
intended to imply recommendation or endorsement by the National Institute of Standards and Technology, nor
is it intended to imply that the products identified are necessarily the best available for the purpose.
\else
  \section*{Acknowledgment}
\fi

\ifCLASSOPTIONcaptionsoff
  \newpage
\fi

\bibliographystyle{IEEEtran}
\bibliography{bivariate}
%



%

\begin{IEEEbiography}[{\includegraphics[width=1in,height=1.25in,clip,keepaspectratio]{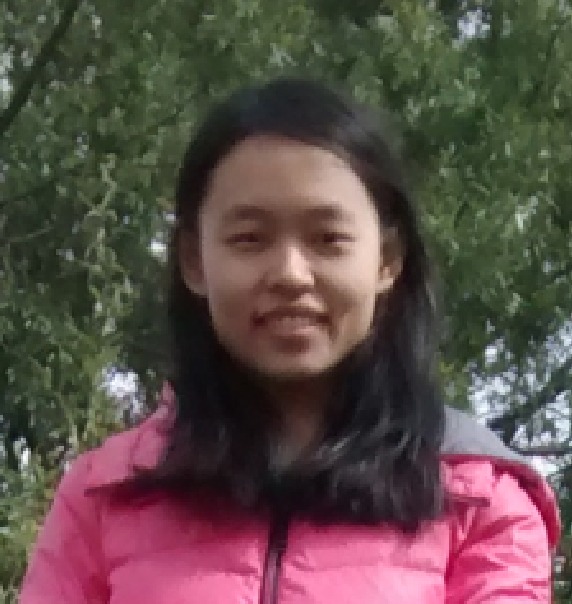}}]{Henan Zhao}
	received the B.E. degree in Computer Science and Information Security from Nankai University, China. She is a Ph.D. student in Computer Science and Electrical Engineering at University of Maryland, Baltimore County. Her research interests include design and evaluation of visualization techniques.
\end{IEEEbiography}

\begin{IEEEbiography}[{\includegraphics[width=1in,height=1.25in,clip,keepaspectratio]{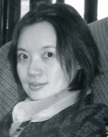}}]{Jian Chen}
received her PhD degree in Computer Science from Virginia Polytechnic Institute and State University (Virginia Tech). She did her postdoctoral work in Computer Science and BioMed at Brown University. She is an Associate Professor in
the Department of Computer Science and Engineering at The Ohio State University, where she directs the 
Interactive Visual Computing Lab. She is also affiliated with the Ohio Translational Data Analytics Institute. 
Her research interests include design and evaluation of visualization techniques, 3D interface, and immersive analytics. 
She is a member of the IEEE and the IEEE Computer Society.
\end{IEEEbiography}




\end{document}

%% file: exp1design.tex
\begin{table}[!t]
\caption{
Experiment I design: 20 participants are assigned to one of the five blocks and use all five bivariate pairs. 
Here, $L_yL_y$: $length_y$-$length_y$ (splitVectors), $L_yL_x$: $length_y$-$length_x$, $LC$: $length$-$color$, $LT$: $length$-$texture$,
and $LCL$: $length_y$-$color/length_x$.
}
\label{tab:experimentdesign}
\scriptsize
\begin{center}
\begin{tabular}{ l l l }
   \toprule
 \centering Block & Participant & Feature-pair   \\
 \midrule
 	1 & P1, P6, P11, P16 & splitVectors, $L_yL_x$, $LC$, $LT$, $LCL$\\
    2 & P2, P7, P12, P17 & $L_yL_x$, $LC$, $LCL$, splitVectors, $LT$\\
    3 & P3, P8, P13, P18 & $LC$, $LCL$, $LT$, splitVectors, $L_yL_x$\\
    4 & P4, P9, P14, P19 & $LT$, $L_yL_x$, splitVectors, $LCL$, $LC$\\
    5 & P5, P10, P15, P20 & $LCL$, $LT$, $L_yL_x$, $LC$, splitVectors\\ 
  \bottomrule
\end{tabular}
\end{center}
\end{table}

%% file: Exp1timeError.tex
\begin{table}[!tp]
	\caption{Summary statistics by tasks. 
    The significant main effects and the high
	effect size (ES) are in \textbf{bold} (none in these observations) and the medium effect size is in \textit{italic}. Effect size is 
eta-square 
labeled ``small'' $(0.01-0.06)$, ``medium'' $[0.06-0.14)$, and ``large'' $\geq 0.14$ effects following Cohen~\cite{cohen1988statistical}.	Post-hoc Tukey grouping results are reported for significant main effects, where $>$ means statistically significantly better and enclosing parentheses mean they belong to the same Tukey group.
}
	\label{tab:new_glm}
	\scriptsize
	\begin{center}
		\begin{tabular}{l l l l}
		\toprule
			\centering Task & Variables & Significance & ES\\
			\midrule
	MAG &  time  & \textbf{F$_{(4,\,384)}$ = 6.8, \textit{p} $<$ 0.0001} & \textit{0.07} \\ 
	    & & \textbf{(LC, LT, LCL, splitVectors)} $>$ \textbf{L$_y$L$_x$} &\\
			& relative error & F$_{(4,\,384)}$ = 0.9, \textit{p} = 0.46 & 0.01 \\ 
			\midrule
	RATIO  &  time  &  \textbf{F$_{(4,\,395)}$ = 6.2,  \textit{p} $<$ 0.0001} & \textit{0.06} \\ 
	& & Three groups: \textbf{A: LC, splitVectors, LT} &\\
	& & \hspace{1.53cm} \textbf{B: splitVectors, LT, LCL} & \\
	& & \hspace{1.53cm} \textbf{C: LT, LCL, L$_y$L$_x$}& \\
	& relative error & F$_{(4,\,395)}$ = 0.8, \textit{p} =  0.50 & 0.01 \\ 
	\midrule
	COMP  &  time  &  \textbf{F$_{(4,\,395)}$ = 10.4, \textit{p} $<$ 0.0001} & \textit{0.09}\\ 
	& & Three groups: \textbf{A: LCL, LC, LT}& \\
	& & \hspace{1.53cm} \textbf{B: LC, splitVectors} & \\
	& & \hspace{1.53cm} \textbf{C: splitVectors, L$_y$L$_x$} & \\
	& accuracy &  ${\chi}^2$ = 0.4, \textit{p} = 0.98 & 0.03\\
	
			\bottomrule
		\end{tabular}
	\end{center}
\end{table}

\begin{figure*}[!thp]
\centering
	\begin{subfigure}{0.33\textwidth}
    \includegraphics[width=\textwidth]{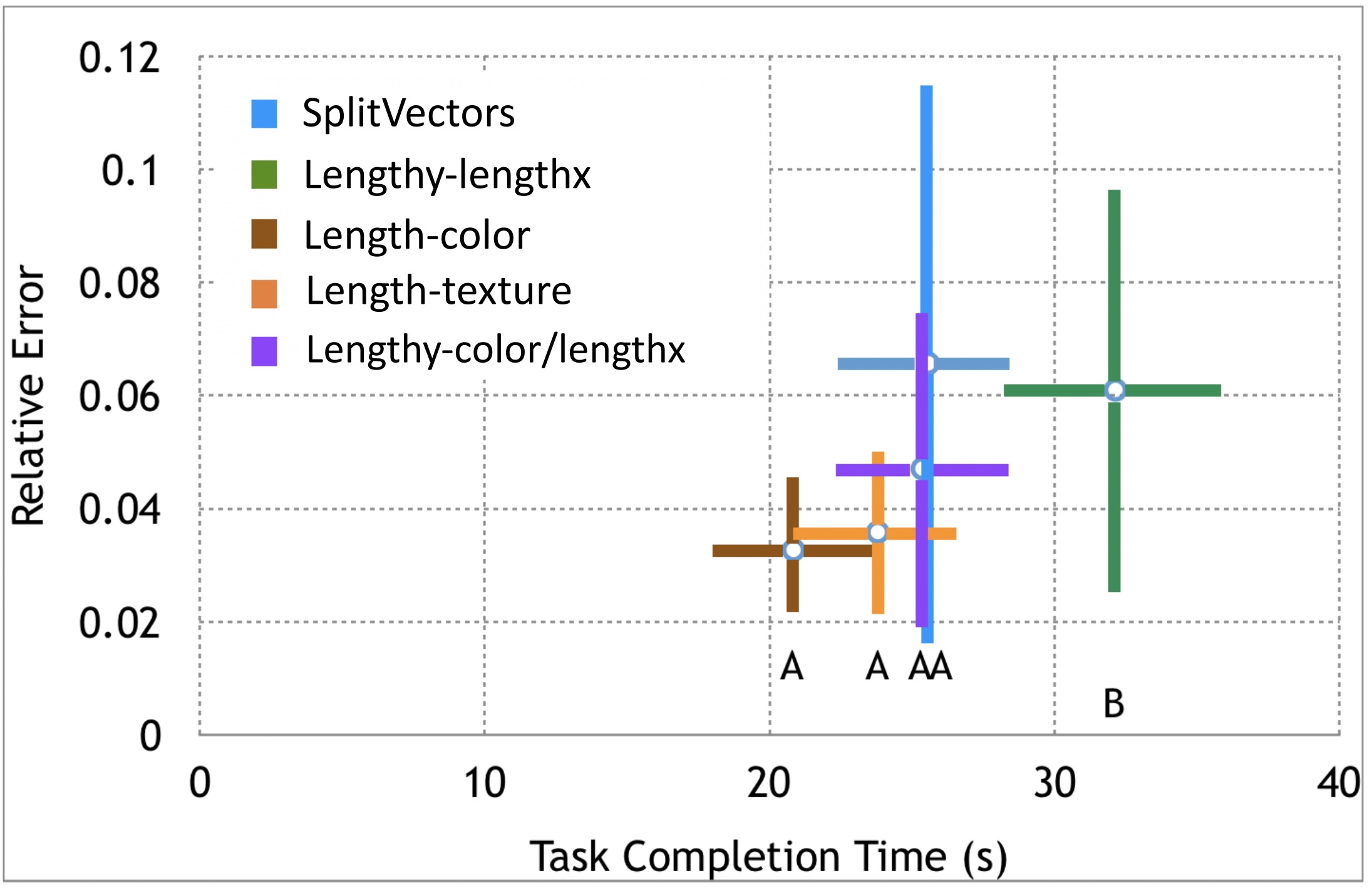}
		\caption{Task 1 (MAG)}
		\label{fig:task1timeError}
	\end{subfigure}
\begin{subfigure}{0.33\textwidth}
		\centering
	\includegraphics[width=\textwidth]{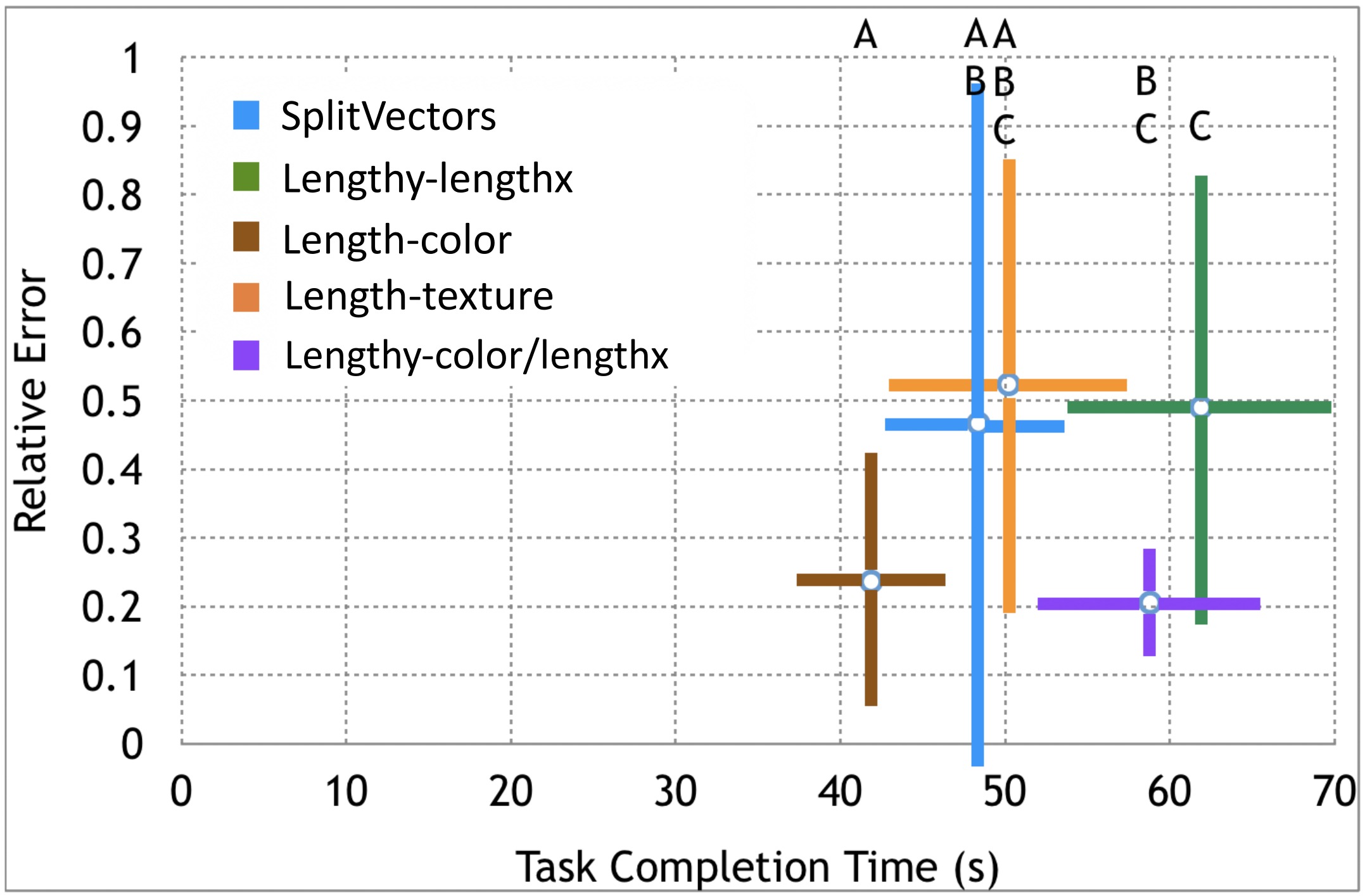}
		\caption{Task 2 (RATIO)}
		\label{fig:task2timeError}
	\end{subfigure}
	\begin{subfigure}{0.33\textwidth}
	\includegraphics[width=\textwidth]{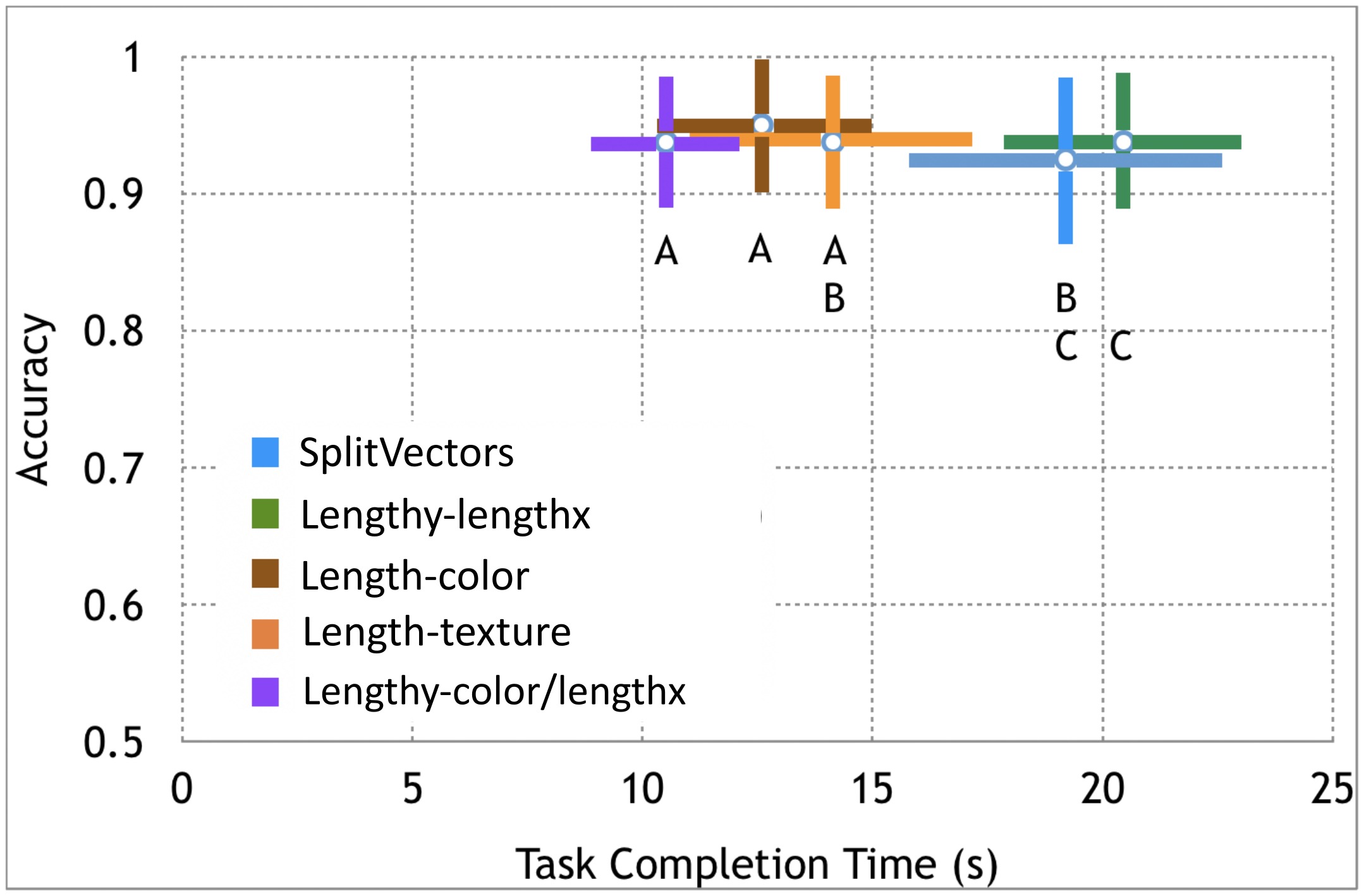}
		\caption{Task 3 (COMP)}
		\label{fig:task3timeError}
	\end{subfigure}

\caption{Task completion time \remove{($log_e$)} and relative error or accuracy by tasks. 
The horizontal axis represents the \remove{$log_e(time)$} mean task completion time while the vertical axis showing the accuracy or relative error. Same letters represent the same post-hoc analysis group. Colors label the feature-pair types. All error bars represent $95\%$ confidence interval.
}

\label{fig:timeError}
\end{figure*}

%% file: figureCase2.tex
\begin{figure*}[!tb]
\centering
\begin{subfigure}{0.45\textwidth}
		\includegraphics[width=\textwidth]{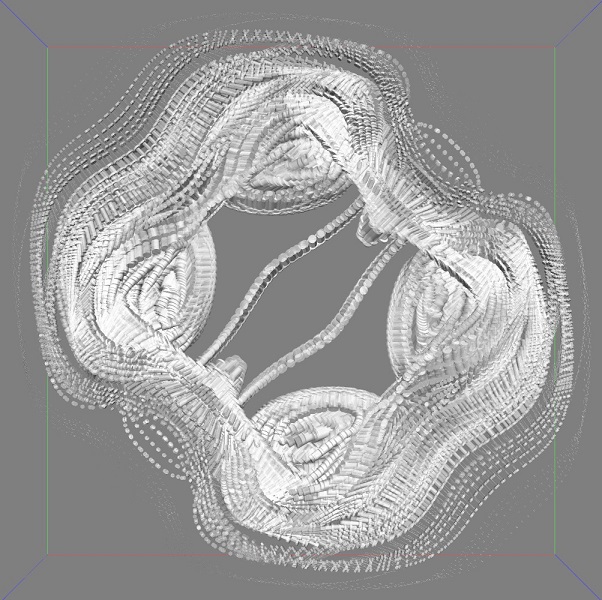}
		\caption{SplitVectors}
		\label{fig:case2LLO}
\end{subfigure} 
\begin{subfigure}{0.45\textwidth}
		\includegraphics[width=\textwidth]{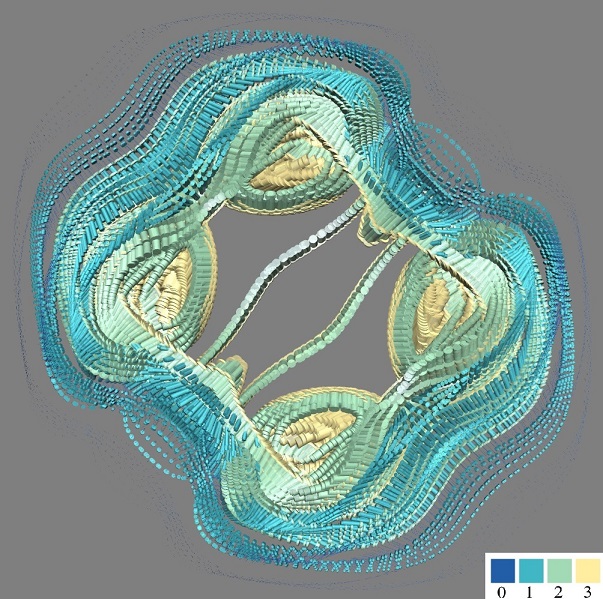}
		\caption{$Length_y$-$color/length_x$}
		\label{fig:case2LCL}
\end{subfigure} 
\begin{subfigure}{0.45\textwidth}
		\includegraphics[width=\textwidth]{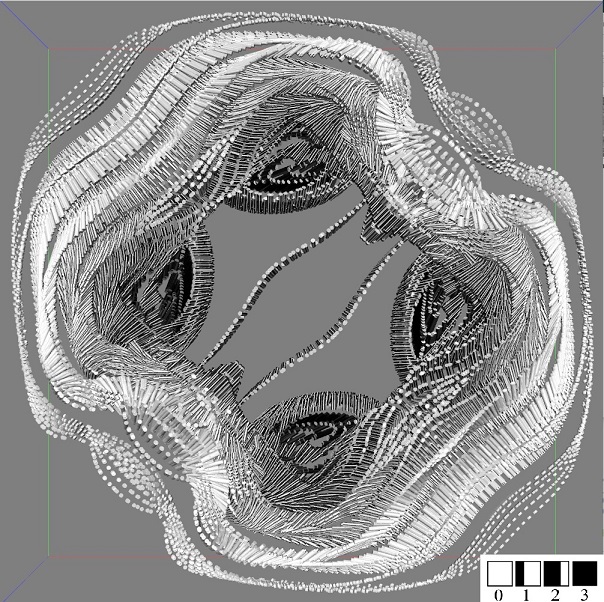}
		\caption{$Length_y$-$texture$}
		\label{fig:case2LT}
\end{subfigure} 
\begin{subfigure}{0.45\textwidth}
		\includegraphics[width=\textwidth]{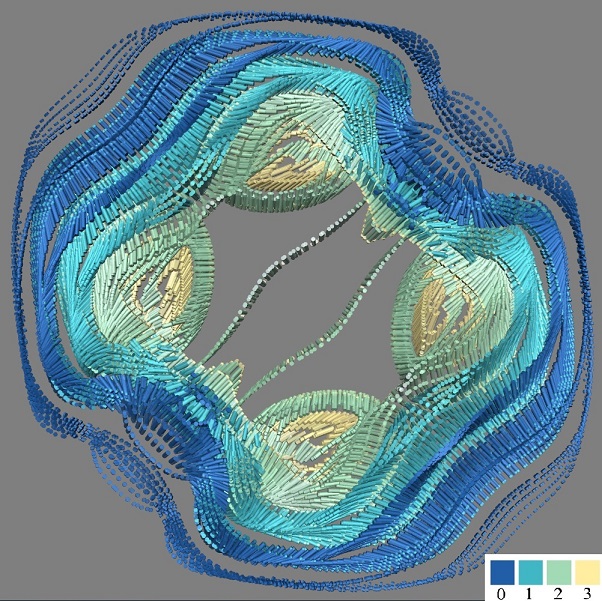}
		\caption{$Length_y$-$color$}
		\label{fig:case2LC}
\end{subfigure} 
\caption{
Contours of simulation data. Size from this viewpoint can guide visual grouping and size in 3D must take advantage of knowledge of the layout of the scene~\cite{eckstein2017humans}. }
\label{fig:case2}
\end{figure*}